%% file: main.tex
\documentclass[%
 reprint,
superscriptaddress,
twocolumn,
 nofootinbib,
 amsmath,
 amssymb,
 aps,
 prx,
]{revtex4-2}
\pdfoutput=1

\usepackage[utf8]{inputenc}
\usepackage[english]{babel}
\usepackage[a4paper, margin=1in]{geometry}
\usepackage{graphicx}
\usepackage{amsmath}
\usepackage{amsfonts}
\usepackage{physics}
\usepackage{bm}
\usepackage[dvipsnames]{xcolor}
\usepackage{tikz}
\usetikzlibrary{matrix}
\usepackage{natbib}
\usepackage[colorlinks=true,linkcolor=teal,citecolor=Maroon,urlcolor=Maroon]{hyperref}
\setcitestyle{numbers, sort&compress}

\usepackage{booktabs, amsthm}
\usepackage{stmaryrd}
\usepackage{makecell}
\usepackage{cleveref}
\crefformat{equation}{(#2#1#3)} %
\usepackage[noend,linesnumbered, ruled, vlined]{algorithm2e}
\renewcommand{\vec}[1]{\mathbf{#1}}
\newcommand{\secref}[1]{Section~\hyperref[#1]{\ref*{#1}}}
\newcommand{\appref}[1]{Appendix~\hyperref[#1]{\ref*{#1}}}
\newcommand{\tabref}[1]{Table~\hyperref[#1]{\ref*{#1}}}
\newcommand{\figref}[1]{fig.~\hyperref[#1]{\ref*{#1}}}
\newcommand{\sfigref}[2]{fig.~\hyperref[#1]{\ref*{#1}(#2)}}

\definecolor{persianpink}{rgb}{0.97,0.5,0.75}

\definecolor{C0}{HTML}{C22D1E}
\definecolor{C1}{HTML}{089B9E}
\definecolor{C2}{HTML}{F6B34B}
\definecolor{C3}{HTML}{67B86F}
\definecolor{C4}{HTML}{CB89BB}

\theoremstyle{definition}
\newtheorem{definition}{Definition}

\begin{document}

\title{%
High-threshold, low-overhead and single-shot decodable fault-tolerant quantum memory}
\author{Thomas R. Scruby}
\email{t.r.scruby@gmail.com}
\affiliation{Okinawa Institute of Science and Technology, Japan}
\author{Timo Hillmann}
\email{timo.hillmann@rwth-aachen.de}
\affiliation{Chalmers University of Technology, Gothenburg, Sweden}
\affiliation{School of Physics,
University of Sydney, Sydney, NSW 2006, Australia}
\author{Joschka Roffe}
\email{joschka@roffe.eu}
\affiliation{University of Edinburgh, United Kingdom}
\affiliation{Freie Universität Berlin, Germany}

\begin{abstract} %
    We present a new family of quantum low-density parity-check codes, which we call radial codes, obtained from the lifted product of a specific subset of classical quasi-cyclic codes. The codes are defined using a pair of integers $(r,s)$ and have parameters $[\![2r^2s,2(r-1)^2,\leq2s]\!]$, with numerical studies suggesting average-case distance linear in $s$. In simulations of circuit-level noise, we observe comparable error suppression to surface codes of similar distance while using approximately five times fewer physical qubits. This is true even when radial codes are decoded using a single-shot approach, which can allow for faster logical clock speeds and reduced decoding complexity. We describe an intuitive visual representation, canonical basis of logical operators and optimal-length stabiliser measurement circuits for these codes, and argue that their error correction capabilities, tunable parameters and small size make them promising candidates for implementation on near-term quantum devices.
\end{abstract}

\maketitle

\section{Introduction}
\label{section:introduction}
\input{intro.tex}

\section{Classical Radial Codes}
\label{section:classical_radial_codes}
\input{classical.tex}

\section{Quantum Radial Codes}
\label{section:quantum_radial_codes}
\input{quantum.tex}

\section{Code Distances of Quantum Radial Codes}
\label{section:distance}
\input{distance.tex}
\input{confinement.tex}

\section{Single-Shot Decoding Under Circuit-Level Noise}
\label{section:Decoding}
\input{decoding.tex}

\section{Conclusion}
\label{section:conclusion}
\input{conclusion.tex}

\section*{Acknowledgements}
T. R. S. acknowledges support from the JST Moonshot R\&D Grant [grant number JPMJMS2061]. 
T. H. acknowledges financial support from the Chalmers Excellence Initiative Nano and the Knut and Alice Wallenberg Foundation through the Wallenberg Centre for Quantum Technology (WACQT). JR is funded by an EPSRC Quantum Career Acceleration Fellowship (grant code: UKRI1224). JR further acknowledges support from EPSRC (EP/T001062/1), EPSRC (EP/X026167/1), BMBF (RealistiQ), and DFG (CRC 183). 
Numerical results presented in this work were obtained using the HPC resources provided by the Scientific Computing and Data Analysis section of the Research Support Division at OIST, and also using resources at the Chalmers Centre for Computational Science and Engineering (C3SE) under project 2024/1-9.
The decoding part of this project resulted from discussions between T. R. S. and T. H. at the 2024 YITP Quantum Error Correction workshop in Kyoto. The authors also acknowledge valuable discussions with Armanda Quintavalle, Boren Gu, Oscar Higgott, Mike Vasmer and Jens Eisert.
The title of this paper is intended as a friendly joke and the authors hope it is received as such.

\input{main.bbl}
\clearpage

\onecolumngrid

\section*{Appendices}

\appendix

\section{Binary PCMs for a small QRC}
\label{app:binary_matrices}
\input{app_pcms.tex}

\section{Strategies for preventing low-distance logicals}
\label{app:distance}
\input{app_lowdistance.tex}

\section{Accuracy of numerically estimated code distances}
\label{app:accuracy}
\input{app_accuracy.tex}

\section{Definitions of codes used in numerics}
\label{app:code_defs}
\input{app_code_defs.tex}

\section{Decoder Optimizations}
\label{app:decoder_optimization}

\input{app_decoder_optimization.tex}

\end{document}

%% file: intro.tex
A quantum error correcting code (QECC) is a strategy for redundantly encoding quantum information in a way that protects this information from environmental noise. The most commonly studied encodings use the collective state of a large number of qubits ($n$) to encode the state of a smaller number of qubits ($k$) so that one encoded state cannot be transformed into another without interacting with at least $d$ of these qubits~\cite{nielsen_quantum_2011}. 
Due to the extreme fragility of quantum systems, QECCs have immense potential value in information-processing tasks, 
but their inherent inefficiency can also lead to significantly increased resource costs. 
In recent years much progress has been made in reducing these overheads, most notably with the development of so-called ``good'' quantum low-density parity-check (LDPC) codes~\cite{Breuckmann_2021,panteleev_asymptotically_2022, dinur_good_2022, leverrier_quantum_2022} which have $k$ and $d$ both scaling linearly with $n$ while also possessing a description in terms of sparse matrices (a desirable property for a number of reasons). 
While these codes require numbers of physical qubits well beyond the scale of current or near-term quantum computing devices, the techniques developed as part of their construction (in particular, the \textit{lifted product} of classical codes~\cite{panteleev_quantum_2022}) have been observed also to be highly effective at creating efficient, small-scale codes~\cite{Panteleev2021degeneratequantum,Roffe2023biastailoredquantum, Raveendran2022finiterateqldpcgkp, xu_constant-overhead_2024,sabo_weight_2024}. 

In this work, we describe a new family of quantum codes, which we call quantum radial codes, constructed using the lifted product. 
The codes have parameters comparable to the bivariate bicycle codes of~\cite{bravyi_high-threshold_2024} while also showing evidence of single-shot decodability under circuit-level noise. 
Additionally, they have an intuitive visual representation, a canonical basis of logical operators and optimal-length circuits for stabiliser measurement. 
Each code is defined using a pair of integers $(r,s)$ and has parameters $[\![n,k,d]\!] = [\![2r^2s,2(r-1)^2,\leq2s]\!]$ and checks of weight $2r$. 
The independent tunability of $k$ and $d$ makes these codes highly flexible and potentially suited for a variety of applications.
\Cref{tab:code_params} summarises properties of the specific instances studied in this work, which have check weights of at most eight. 

\begin{table}[!b]
    \centering
    \begin{tabular}{lllll} \hline
       $\llbracket n, k, d\rrbracket$ & \makecell{Net \\ Encoding \\ Rate} & \makecell{Pseudo \\ Threshold} & $p_Z(10^{-3})$ \\ \hline \hline 
       $\llbracket 90, 8, 10\rrbracket$  & 2/45 & $0.4 \%$ & $\approx 6 \times 10^{-6}$ \\
       $\llbracket 352, 18, 20\rrbracket$ & $\approx 1/40$ & $0.6 \%$ & $\approx 3 \times 10^{-8}$ \\ 
    \end{tabular} 
    \caption{Code parameters, thresholds, and $Z$ word error rates at physical error rate $p = 10^{-3}$ for the quantum radial codes studied in this work. 
    The net encoding rate counts both data qubits and ancilla qubits used for syndrome readout and for our codes is given by $k / 2n$.
    The value of the $Z$ word error rate for the $\llbracket 352, 18, 20\rrbracket$ is obtained through a fit, see \sfigref{fig:main_results}{b}.}
    \label{tab:code_params}
\end{table}

The rest of this manuscript is structured as follows. In \cref{section:classical_radial_codes} we describe the classical codes used as input to the lifted product and then in \cref{section:quantum_radial_codes} we study the quantum codes produced. In \cref{section:distance} we numerically investigate the distances of these codes and discuss methods for ensuring higher average-case distance. Finally, in \cref{section:Decoding} we describe our techniques for simulating the performance of these codes under circuit-level noise and present evidence for their single-shot decodability in this setting. The scripts used to obtain all the numerical results in this work can be found open-source at~\cite{github_repo}.

%% file: classical.tex
We begin by defining what we call \textit{classical radial codes}, which are a subset of the quasi-cyclic codes presented in~\cite{fossorier_quasicyclic_2004}. These codes are defined using a pair of integers $(r,s)$ and have parity check matrices of the form

\begin{equation}
    H = \begin{pmatrix}
        \lambda_s^{a_{(0,0)}}   & \lambda_s^{a_{(0,1)}}   & . & . & . & \lambda_s^{a_{(0,r-1)}}   \\
        \lambda_s^{a_{(1,0)}}   & \lambda_s^{a_{(1,1)}}   & . & . & . & \lambda_s^{a_{(1,r-1)}}   \\
        .                       & .                       & . &   &   & .                         \\
        .                       & .                       &   & . &   & .                         \\
        .                       & .                       &   &   & . & .                         \\
        \lambda_s^{a_{(r-1,0)}} & \lambda_s^{a_{(r-1,1)}} & . & . & . & \lambda_s^{a_{(r-1,r-1)}} \\
    \end{pmatrix}
    \label{eq:classical_H}
\end{equation}

\noindent where $\lambda_s^a$ is an element of the ring of circulants of length $s$. $\lambda_s^a$ has a binary matrix representation as the $s \times s$ identity matrix with each row shifted to the right by $a$ places, which we write as $\mathfrak{B}(\lambda_s^a)$. We will also use $\mathfrak{B}(H)$ to mean the matrix obtained by replacing each element of $H$ with its binary representative. The code can also be described more concisely using just a matrix

\begin{equation}
    A = \begin{pmatrix}
        a_{(0,0)}   & a_{(0,1)}    & . & . & . & a_{(0,r-1)}   \\
        a_{(1,0)}   & a_{(1,1)}    & . & . & . & a_{(1,r-1)}   \\
        .           & .            & . &   &   & .             \\
        .           & .            &   & . &   & .             \\
        .           & .            &   &   & . & .             \\
        a_{(r-1,0)} & a_{(r-1,1)}  & . & . & . & a_{(r-1,r-1)} \\
    \end{pmatrix}_s
\end{equation}
It is known that provided the elements of $A$ satisfy
\begin{equation}
    \label{eq:square_condition}
    a_{(u_1,u_1')} - a_{(u_1,u_2')} - a_{(u_2,u_1')} + a_{(u_2,u_2'}) \neq 0 \mod s,
\end{equation}
the girth (minimum cycle length) in the Tanner graph of the associated code is at least six~\cite{fossorier_quasicyclic_2004}. In addition, we will require that

\begin{enumerate}
    \item $s$ is prime
    \item $r \leq s$
    \item the rows of $A$ are linearly independent
\end{enumerate}
The motivation for these restrictions will be explained shortly, but first, it will be helpful to introduce a visual description of the structure of these codes. 

For any code with parity check matrix of the form in \cref{eq:classical_H} we can arrange the bits and checks in a pattern of $r$ concentric rings containing $s$ spokes. This is done by enumerating the rings and spokes and then assigning to every row $i$ of $\mathfrak{B}(H)$ a coordinate $(u,v)$ where
\begin{equation}
    \begin{split}
    u = \textrm{ring number} &= \lfloor i/s \rfloor \\
    v = \textrm{spoke number} &= i \mod s
    \end{split}
\end{equation}

\noindent (where $\lfloor . \rfloor$ is the floor operator) and similarly for the columns. 
This gives us one bit and one check for each independent coordinate $(0 \leq x < r, 0 \leq y < s)$. 
Each ring then contains $s$ bits and $s$ checks (one pair from each spoke) and each spoke contains $r$ bits and $r$ checks (one pair from each ring). 
When arranged in this way each check in the code contains exactly one bit from each ring and each bit is part of exactly one check from each ring. 
For clarity, we will always use $u$ to refer to ring indices and $v$ to refer to spoke indices. 
Additionally, we will use primed indices for bit coordinates and unprimed indices for check coordinates.
\begin{figure}[!b]
    \includegraphics[width=0.23\textwidth]{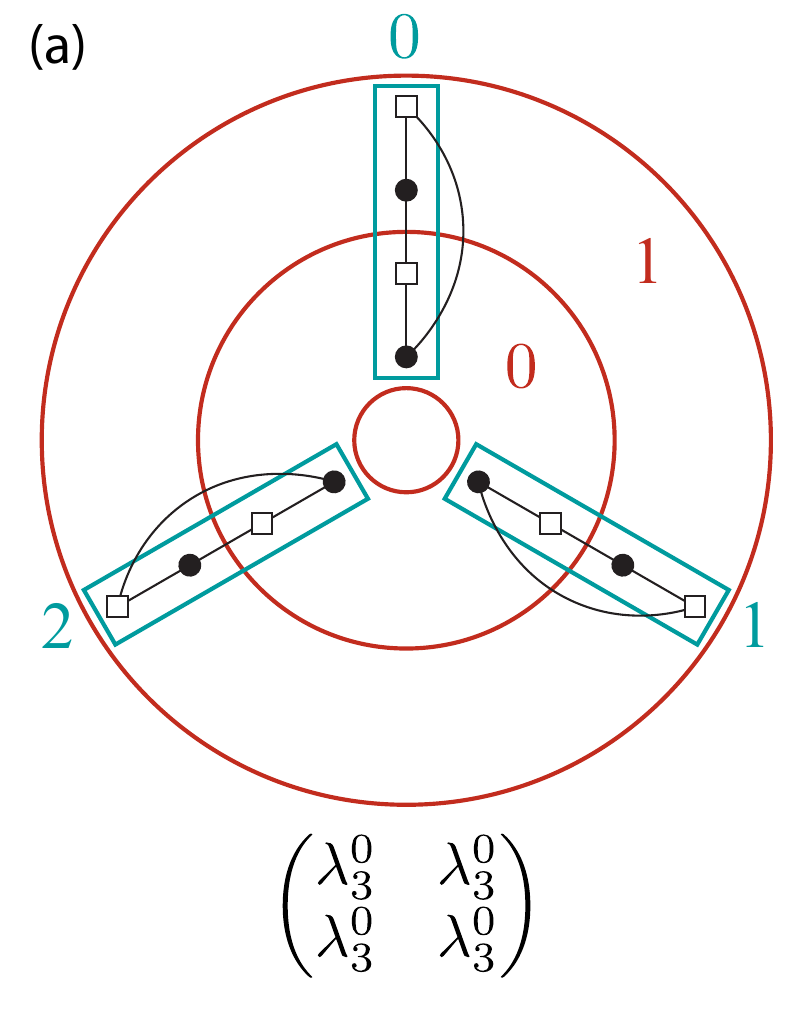}
    \includegraphics[width=0.23\textwidth]
    {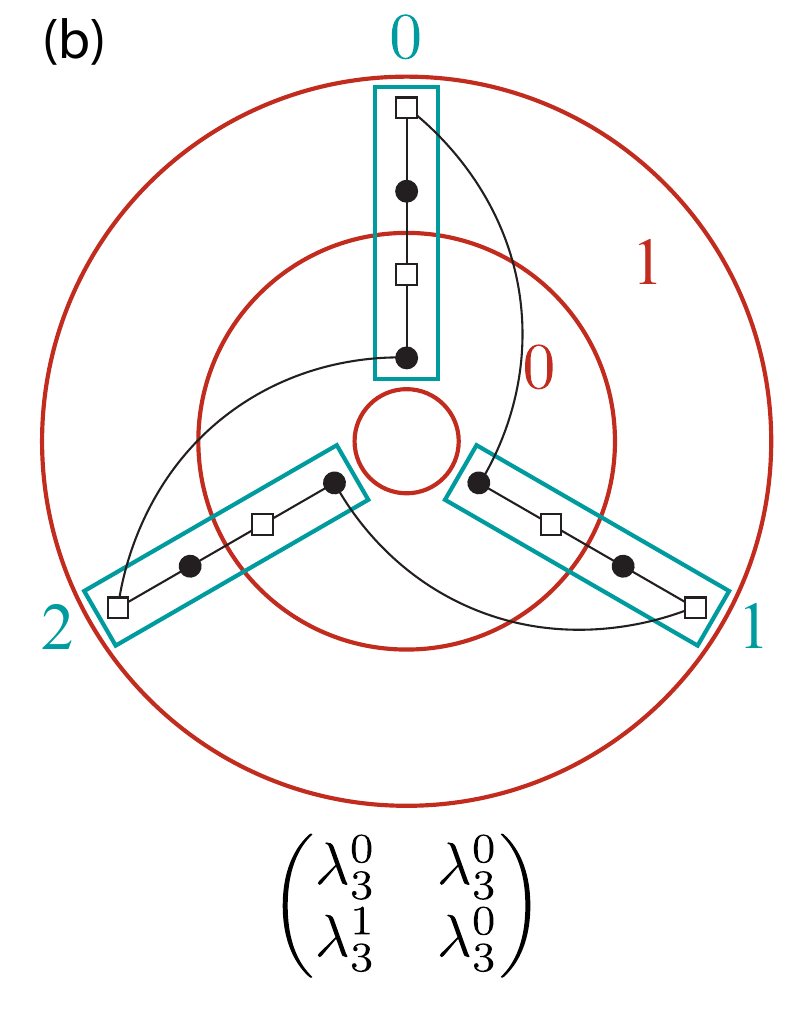}
    \caption{Radial layouts for two examples of $(r,s)=(2,3)$ classical codes with parity check matrices given in \cref{eq:classical_example_a} and \cref{eq:classical_example_b} respectively. Bits/checks are shown as circles/squares. Rings (and ring indices) are shown in red. Spokes (and spoke indices) are shown in blue.}
    \label{fig:small_radial_examples}
\end{figure}
Two examples are shown in \cref{fig:small_radial_examples}, that is, in \sfigref{fig:small_radial_examples}{a} we have a code

\noindent\begin{minipage}{0.15\textwidth}
    \begin{equation*}
        \begin{split}
            A =& 
            \begin{pmatrix}
                0 & 0 \\
                0 & 0 
            \end{pmatrix}_3 \\
            \\
            H =& 
            \begin{pmatrix}
                \lambda_3^0 & \lambda_3^0 \\
                \lambda_3^0 & \lambda_3^0 
            \end{pmatrix}
        \end{split}
    \end{equation*}
\end{minipage}
~
\begin{minipage}{0.33\textwidth}
    \begin{equation}
        \mathfrak{B}(H)
          = \begin{pmatrix}
            1 & 0 & 0 & 1 & 0 & 0 \\
            0 & 1 & 0 & 0 & 1 & 0 \\
            0 & 0 & 1 & 0 & 0 & 1 \\
            1 & 0 & 0 & 1 & 0 & 0 \\
            0 & 1 & 0 & 0 & 1 & 0 \\
            0 & 0 & 1 & 0 & 0 & 1 
        \end{pmatrix}
      \label{eq:classical_example_a}
    \end{equation}
    \vspace{0.05pt}
\end{minipage}
where each spoke is a copy of the 2-bit cyclic repetition code, while in \sfigref{fig:small_radial_examples}{b} we have a code

\noindent\begin{minipage}{0.15\textwidth}
    \begin{equation*}
        \begin{split}
            A =& 
            \begin{pmatrix}
                0 & 0 \\
                1 & 0 
            \end{pmatrix}_3 \\
            \\
            H =& 
            \begin{pmatrix}
                \lambda_3^0 & \lambda_3^0 \\
                \lambda_3^1 & \lambda_3^0 
            \end{pmatrix}
        \end{split}
    \end{equation*}
\end{minipage}
~
\begin{minipage}{0.33\textwidth}
    \begin{equation}
        \mathfrak{B}(H)
          = \begin{pmatrix}
            1 & 0 & 0 & 1 & 0 & 0 \\
            0 & 1 & 0 & 0 & 1 & 0 \\
            0 & 0 & 1 & 0 & 0 & 1 \\
            0 & 1 & 0 & 1 & 0 & 0 \\
            0 & 0 & 1 & 0 & 1 & 0 \\
            1 & 0 & 0 & 0 & 0 & 1
        \end{pmatrix}
      \label{eq:classical_example_b}
    \end{equation}
    \vspace{0.05pt}
\end{minipage}
with connections between the spokes, resulting in a 6-bit cyclic repetition code.

By examining the parity check matrices of these codes more closely we can see that the values $a_{u,u'}$ can be interpreted as a description of the connections between checks of ring $u$ and bits of ring $u'$. Specifically, if $a_{u,u'} = x$ then each check of ring $u$ is connected to the bit of ring $u'$ that is $x$ spokes ahead of it. When all $a_{u,u'}=0$ (as in \sfigref{fig:small_radial_examples}{a}) each check connects only to bits in its own spoke. Note that because the values of $a_{u,u'}$ depend only on the ring coordinates of the relevant checks and bits this representation of the Tanner graph is rotationally symmetric with order $s$. 

We can now use this perspective to understand the significance of the conditions described above, starting with a derivation of Eq.~\cref{eq:square_condition}.
Consider a bit with coordinates $(u_1',v_1')$ and two checks connected to this bit, which must have coordinates $(u_1,v_1' - a_{u_1,u_1'})$ and $(u_2,v_1' - a_{u_2,u_1'})$, with addition implicitly performed $\mod s$. Under what circumstances can these checks also share a second bit, i.e., resulting in a length-4 cycle?
The first check is connected to bits with coordinates $(u_2',(v_1'-a_{(u_1,u_1')})+a_{(u_1,u_2')})$ and the second to bits with coordinates $(u_3',(v_1'-a_{(u_2,u_1')})+a_{(u_2,u_3')})$. 
For two of these bits to be the same we need the ring coordinates to be equal, $u_2' = u_3'$, and also the spoke coordinates to be equal
\begin{equation}
    v_1' - a_{(u_1,u_1')} + a_{(u_1,u_2')} = v_1' - a_{(u_2,u_1')} + a_{(u_2,u_2')}.
\end{equation}
By requiring that 
this condition is not met for any $u_1,u_2,u_1',u_2'$ we obtain \cref{eq:square_condition} and guarantee a certain amount of expansion in the Tanner graph, since checks can only share at most one bit. To see an example, compare the Tanner graphs of \sfigref{fig:small_radial_examples}{a}, where $A$ does not satisfy \cref{eq:square_condition}, and \sfigref{fig:small_radial_examples}{b}, where it does.  
With this in mind, the other three restrictions have the following effects.

\textit{$s$ is prime:} As mentioned above, the radial representation of the Tanner graph is rotationally symmetric with order $s$, and so any disjoint subgraphs contained in the graph must also have rotational symmetry with order that is some factor of $s$. By choosing $s$ to be prime we thus ensure that all rotational symmetries of the graph are of order $s$ or $1$, and when combined with \cref{eq:square_condition} this ensures that the Tanner graph consists of a single connected component. This is because a disjoint subgraph with order $s$ symmetry (that is not just the entire graph) must contain every bit and check in some subset of the rings and no bits or checks from other rings, but because each bit/check is connected to one check/bit from each ring such a subgraph is not possible.
On the other hand, if there is a disjoint subgraph with order $1$ symmetry then there must be $s$ disjoint copies of this subgraph which are mapped onto each other by the order $s$ symmetry of the total graph.
Each of these subgraphs can then contain at most $r$ unique bit nodes (as there are only $rs$ bits in the code), but each bit is connected to $r$ checks and by \cref{eq:square_condition} these checks are connected to $r(r-1)$ unique bits so there can be no disjoint subgraphs of this size. The only remaining option is for there to be no nontrivial disjoint subgraphs.

\textit{$r \leq s$:} This is necessary because the $r$ checks which share a given bit cannot have any other shared bits and must also be connected to one bit from each ring, so there must be at least $r$ unique bits in each ring.
We can also see that the distance of the code must be at least $r$, as any nonzero bit of a given codeword is connected to $r$ checks and each of these must be connected to at least one other nonzero bit of this codeword for the checks to be satisfied.

\textit{The rows of $A$ are linearly independent:} Given any row $H_i$ of $H$, the sum of all the rows of $\mathfrak{B}(H_i)$ is the all-$1$s vector. This means that there is one linear dependency in $\mathfrak{B}(H)$ for each pair of rows in $H$, giving $r-1$ linear dependencies. If we then choose the rows of $A$ to be linearly independent then we find that with high probability\footnote{We thank Min Ye for this observation and for identifying an issue with a related claim in the previous version of this work. We note also that the instances where this is not the case are typically outside the parameter regime we will consider, although it is an interesting question for future work to understand the origin of these instances and identify a better set of conditions on the input matrices. Practically, we verify this condition on the number of linear dependencies explicitly for all generated codes.} these are the only linear dependencies. We then have $rs$ bits and $rs - (r-1)$ independent checks, giving $(r-1)$ encoded bits of information.

Finally, we can choose a basis for the codewords of these codes and show that the distance is $2s$. Enumerating the encoded bits from $0$ to $r-2$, let the $i^\textrm{th}$ codeword correspond to setting all bits in the $i^\textrm{th}$ ring and the $(r-1)^\textrm{th}$ ring to be $1$, and all other bits to be $0$ (this is a codeword because each check in the code acts on exactly one bit from each ring). These $r-1$ codewords are all linearly independent and of weight $2s$, and because they overlap only on the $(r-1)^\textrm{th}$ ring, and this overlap always has size $s$, any linear combination of them always has weight at least $2s$.
These codes then have parameters $[rs,r-1,2s]$, with each node in the Tanner graph having degree $r$.

%% file: quantum.tex
Quantum radial codes (QRCs) are obtained from the lifted product of two classical radial codes. For a (ring-theoretic) ring $R$ and two ring-valued matrices $H_1 \in R^{m_1 \times n_1}$ and $H_2 \in R^{m_2 \times n_2}$ the lifted product is defined as 

\begin{equation}
    \begin{split}
        &H_X = [H_1 \otimes I_{m_2} | I_{m_1} \otimes H_2] \\
        &H_Z = [I_{n_1} \otimes H_2^* | H_1^* \otimes I_{n_2}] 
    \end{split}
\end{equation}
where $H^*$ is the conjugate transpose of $H$, obtained by transposing the matrix and taking the inverse of each element~\cite{panteleev_quantum_2022}.
If $H$ is a classical radial code then $H^*$ will also be a classical radial code (in fact, it is just $H$ with the bits and checks exchanged). 

\subsection{A Small Example}
\label{subsection:small_example}

To understand the structure of QRCs we can start with a simple example, taking (from \sfigref{fig:small_radial_examples}{b})
\begin{equation}
    H_1 = H_2 = H =
    \begin{pmatrix}
        \lambda_3^0 & \lambda_3^0 \\
        \lambda_3^1 & \lambda_3^0
    \end{pmatrix}
\end{equation}
which has conjugate transpose %
\begin{equation}
    H_1^{*} = H_2^* = H^* =
    \begin{pmatrix}
        \lambda_3^0 & \lambda_3^2 \\
        \lambda_3^0 & \lambda_3^0
    \end{pmatrix}\rm.
\end{equation}
so that we obtain a quantum code with parity check matrices
\begin{equation}
\label{eq:example_hx}
H_X = 
\resizebox{.4\textwidth}{!}{
\begin{tikzpicture}[baseline]
    \matrix (M1) [matrix of nodes,{left delimiter=(},{right delimiter=)},nodes={minimum size=7mm},ampersand replacement=\&]
    {
        $\lambda_3^0$ \& 0 \& $\lambda_3^0$ \& 0 \& |[fill=C2]| $\lambda_3^0$ \& |[fill=C2]| $\lambda_3^0$ \& 0 \& 0 \\
        0 \& $\lambda_3^0$ \& 0 \& $\lambda_3^0$ \& |[fill=C2]| $\lambda_3^1$ \& |[fill=C2]| $\lambda_3^0$ \& 0 \& 0 \\
        $\lambda_3^1$ \& 0 \& $\lambda_3^0$ \& 0 \& 0 \& 0 \& |[fill=C3]| $\lambda_3^0$ \& |[fill=C3]| $\lambda_3^0$ \\
        0 \& $\lambda_3^1$ \& 0 \& $\lambda_3^0$ \& 0 \& 0 \& |[fill=C3]| $\lambda_3^1$ \& |[fill=C3]| $\lambda_3^0$ \\
    };
    \draw (M1-1-5.north west) -- (M1-4-4.south east);
    \node [anchor=south] at (M1-1-1.north) {\color{C0} 0}; %
    \node [anchor=south] at (M1-1-2.north) {\color{C0} 1}; %
    \node [anchor=south] at (M1-1-3.north) {\color{C1} 0}; %
    \node [anchor=south] at (M1-1-4.north) {\color{C1} 1}; %
    \node [anchor=south] at (M1-1-5.north) {\color{C2} 0}; %
    \node [anchor=south] at (M1-1-6.north) {\color{C2} 1}; %
    \node [anchor=south] at (M1-1-7.north) {\color{C3} 0}; %
    \node [anchor=south] at (M1-1-8.north) {\color{C3} 1}; %
    \node [anchor=west] at (M1-1-8.east) {\color{C2} \hspace{.5cm}0}; %
    \node [anchor=west] at (M1-2-8.east) {\color{C2} \hspace{.5cm}1}; %
    \node [anchor=west] at (M1-3-8.east) {\color{C3} \hspace{.5cm}0}; %
    \node [anchor=west] at (M1-4-8.east) {\color{C3} \hspace{.5cm}1}; %
\end{tikzpicture}
}
\end{equation}

\begin{equation}
\label{eq:example_hz}
H_Z =
\resizebox{.4\textwidth}{!}{
\begin{tikzpicture}[baseline]
    \matrix (M1) [matrix of nodes,{left delimiter=(},{right delimiter=)},nodes={minimum size=7mm},ampersand replacement=\&]
    {
        |[fill=C0!80]| $\lambda_3^0$ \& |[fill=C0!80]| $\lambda_3^2$ \& 0 \& 0 \& $\lambda_3^0$ \& 0 \& $\lambda_3^2$ \& 0 \\
        |[fill=C0!80]| $\lambda_3^0$ \& |[fill=C0!80]| $\lambda_3^0$ \& 0 \& 0 \& 0 \& $\lambda_3^0$ \& 0 \& $\lambda_3^2$ \\
        0 \& 0 \& |[fill=C1!50]| $\lambda_3^0$ \& |[fill=C1!50]| $\lambda_3^2$ \& $\lambda_3^0$ \& 0 \& $\lambda_3^0$ \& 0 \\
        0 \& 0 \& |[fill=C1!50]| $\lambda_3^0$ \& |[fill=C1!50]| $\lambda_3^0$ \& 0 \& $\lambda_3^0$ \& 0 \& $\lambda_3^0$ \\
    };
    \draw (M1-1-5.north west) -- (M1-4-4.south east);
    \node [anchor=south] at (M1-1-1.north) {\color{C0} 0}; %
    \node [anchor=south] at (M1-1-2.north) {\color{C0} 1}; %
    \node [anchor=south] at (M1-1-3.north) {\color{C1} 0}; %
    \node [anchor=south] at (M1-1-4.north) {\color{C1} 1}; %
    \node [anchor=south] at (M1-1-5.north) {\color{C2} 0}; %
    \node [anchor=south] at (M1-1-6.north) {\color{C2} 1}; %
    \node [anchor=south] at (M1-1-7.north) {\color{C3} 0}; %
    \node [anchor=south] at (M1-1-8.north) {\color{C3} 1}; %
    \node [anchor=west] at (M1-1-8.east) {\color{C0} \hspace{.5cm}0}; %
    \node [anchor=west] at (M1-2-8.east) {\color{C0} \hspace{.5cm}1}; %
    \node [anchor=west] at (M1-3-8.east) {\color{C1} \hspace{.5cm}0}; %
    \node [anchor=west] at (M1-4-8.east) {\color{C1} \hspace{.5cm}1}; %
\end{tikzpicture}
}.
\end{equation}
Readers may find it helpful to refer to the full binary representations of these matrices (\cref{app:binary_matrices}) during the following discussion.

We can think of the quantum code as being built from two copies of $H$ (orange and green) and two copies of $H^*$ (red and blue). We will call the copies of $H$ ``$X$ codes'' and the copies of $H^*$ ``$Z$ codes''. Each qubit is associated with a bit in one of these four codes, while each stabiliser is associated with a check ($X$ stabilisers with the checks of the $X$ codes and $Z$ stabilisers with those of the $Z$ codes). The coloured row/column indices show the associated classical code and ring within that code, of each qubit and stabiliser in the quantum code. 

\begin{figure}
    \includegraphics[width=.5\textwidth]{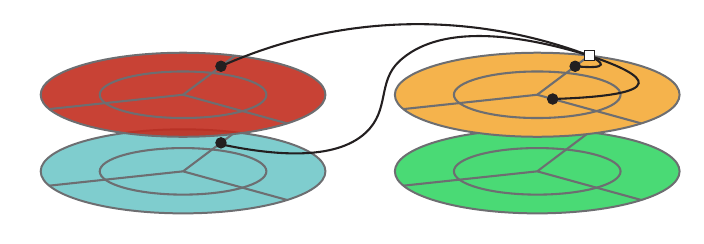}
    \caption{Structure of the QRC with PCMs \cref{eq:example_hx} and \cref{eq:example_hz}. The code is made of four $(r,s) = (2,3)$ classical radial codes, with two (red and blue) providing $Z$ stabilisers and two (orange and green) providing $X$ stabilisers. An example of a $X$ stabiliser from ring $1$ of the orange code (supported on a qubit from each ring of the orange code and a ring $1$ qubit in each of the red and blue codes) is also shown.}
    \label{fig:small_stack}
\end{figure}

The support of each $X$ stabiliser of the quantum code consists of the support of a ring $u$ check in one of the $X$ codes plus one qubit from ring $u$ in each of the $Z$ codes. Importantly, each stabiliser in ring $u$ of a given $X$ code is assigned a \textit{unique} qubit from ring $u$ of each $Z$ code so that the product of all $X$ stabilisers in any ring of an $X$ code is supported on all qubits of that $X$ code plus all ring $u$ qubits in all $Z$ codes. Similarly, each $Z$ stabiliser has the support of a ring $u$ check from a classical $Z$ code plus one qubit from ring $u$ in each of the $X$ codes, and the assignments of these qubits are unique in the same way. All checks have weight $4$, because they act on two $X$ code qubits and two $Z$ code qubits. A representation of this structure is shown in \cref{fig:small_stack}.

To see how many logicals we should expect we can count the number of relations between stabiliser generators. As explained above, the product of all $X$ stabilisers from ring $u$ of the orange code is supported on all qubits of the orange code and all ring $u$ qubits in the red and blue codes. The product of all stabilisers from both rings of the orange code is then supported on all qubits of the red and blue codes and none of the qubits of the orange code. The same is true for the stabilisers of the green code so the product of all $X$ stabilisers is identity on all qubits. The same argument can be made for the $Z$ stabilisers. These are the only relations so we have only two encoded qubits. 

We can also use the radial structure of the code to find a basis for these logical operators. We can see that any classical codeword from one of the $Z$ codes lifts to a logical $X$ operator in the quantum code because if this operator is supported only on, e.g., red qubits then it only intersects $Z$ stabilisers associated with the checks of the red code, and by definition, it commutes with these stabilisers because its support is a codeword of the red code.
We know that a basis for the codewords of a classical radial code is all bits in any pair of adjacent rings, and here we only have two rings so the logical operator is supported on all qubits of the red code.
A similar operator is supported on all qubits of the blue code but these two are equivalent up to composition with stabilisers.
There are also logical $X$ operators supported on all ring $u'$ qubits of the orange and green codes. We can see this because each $Z$ stabiliser acts on one qubit in each of these rings, so it will intersect two qubits of this operator (one orange and one green).
The operators associated with $u'=0$ and $u'=1$ are once again equivalent up to stabilisers. Finally, we can identify a pair of logical $Z$ operators, one supported on the $X$ codes and one on the $Z$ codes, in the same way. This is consistent with the fact that we expected two encoded qubits from counting relations and gives a basis of logical operators with weight $2s = 6$.

\begin{figure}
    \includegraphics[width=.45\textwidth]{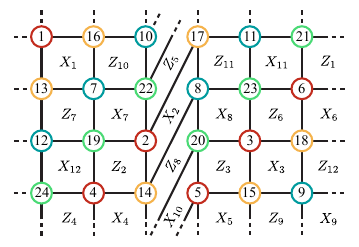}
    \caption{Embedding of the radial code described by \cref{eq:example_hx} and \cref{eq:example_hz} onto a twisted $2$-torus, where it is equivalent to a surface code. Qubits are circles with colouring showing which of the classical radial codes in \cref{fig:small_stack} they belong to. $X$ and $Z$ stabilisers are square faces.}
    \label{fig:toric_code}
\end{figure}

Before generalising these results we observe that this example of a radial code has a different (and more familiar) representation as a surface code on a twisted torus (\cref{fig:toric_code}). Readers who wish to check this explicitly can refer to the binary PCMs in \cref{app:binary_matrices}. This makes it obvious that this code is actually distance $4$ rather than $6$. We will see later that this is not specific to this case, and that while we can generalise the logical basis described here the resulting upper bound on the distance is usually not tight.

\subsection{The General Case}
\label{subsection:general_case}

A general QRC is built from $r$ copies of $H_1$ and $r$ copies of $H_2^*$. Each qubit and stabiliser is assigned a coordinate $(c,u,v)$ where $0 \leq c < 2r$ indexes the code to which it belongs. We will also use the indices $0 \leq x,z < r$ to index $X$ and $Z$ codes separately. 

$H_X$ for a general QRC has the form

\begin{equation}
    H_X = (H_X^{(Z)} | H_X^{(X)})
\end{equation}

\noindent where

\begin{equation}
    \begin{split}
     &H_X^{(Z)}= \\
        &\begin{pmatrix}
            H_1^{(0,0)}I_r   & H_1^{(0,1)}I_r   & . & . & . & H_1^{(0,r-1)}I_r   \\
            H_1^{(1,0)}I_r   & H_1^{(1,1)}I_r   & . & . & . & H_1^{(1,r-1)}I_r   \\
            .                & .                & . &   &   & .                  \\
            .                & .                &   & . &   & .                  \\
            .                & .                &   &   & . & .                  \\
            H_1^{(r-1,0)}I_r & H_1^{(r-1,1)}I_r & . & . & . & H_1^{(r-1,r-1)}I_r \\
        \end{pmatrix}
    \end{split}
\end{equation}

\noindent describes the connections of $X$ stabilisers to qubits of the $Z$ codes ($I_r$ is the $r \times r$ identity matrix) and 

\begin{equation}
        H_X^{(X)} = 
        \begin{pmatrix}
            H_2 & 0     & . & . & . & 0   \\
            0     & H_2 & . & . & . & 0   \\
            .     & .   & . &   &   & .   \\
            .     & .   &   & . &   & .   \\
            .     & .   &   &   & . & .   \\
            0     & 0   & . & . & . & H_2 \\
    \end{pmatrix}
\end{equation}

\noindent describes the connections to qubits of the $X$ codes. In this block form each column of $H_X^{(Z)}$/$H_X^{(X)}$ is associated with all qubits of a $Z$/$X$ code while each row is associated with all $X$ stabilisers of an $X$ code. We can then see that the support of an arbitrary $X$ stabiliser from $X$ code $x$, ring $u$, will be a single qubit from each ring of this $X$ code (the same support as a check of $H_2$) and a single qubit from ring $u$ of each $Z$ code. This second point is due to the fact that each $H_1^{(i,j)}I_r$ can be expanded as 

\begin{equation}
\begin{split}
    H_1^{(i,j)}I_r = &\lambda_s^{a^1_{(i,j)}}I_r = \\
    &\begin{pmatrix}
        \lambda_s^{a^1_{(i.j)}} & 0                       & . & . & . & 0                     \\
        0                       & \lambda_s^{a^1_{(i,j)}} & . & . & . & 0                     \\
        .                       & .                       & . &   &   & .                       \\
        .                       & .                       &   & . &   & .                       \\
        .                       & .                       &   &   & . & .                       \\
        0                       & 0                       & . & . & . & \lambda_s^{a^1_{(i,j)}} \\
    \end{pmatrix}
\end{split}
\end{equation}

\noindent where each row $u$ is associated with all checks from the $u$th ring of $X$ code $x$ and each column $u'$ is associated with all qubits from the $u'$th ring of $Z$ code $z$. Because the matrix is diagonal $X$ checks from ring $u$ can only connect to $Z$ code qubits from ring $u$, and because $\lambda_s^{a^1_{(i,j)}}$ has exactly one non-zero entry in each row and column each $X$ check is connected to exactly one qubit from each $Z$ code, and no two $X$ stabilisers from the same $X$ code $x$ and ring $u$ share a qubit in $Z$ code $z$.

\begin{figure}
   \includegraphics[width=.5\textwidth]{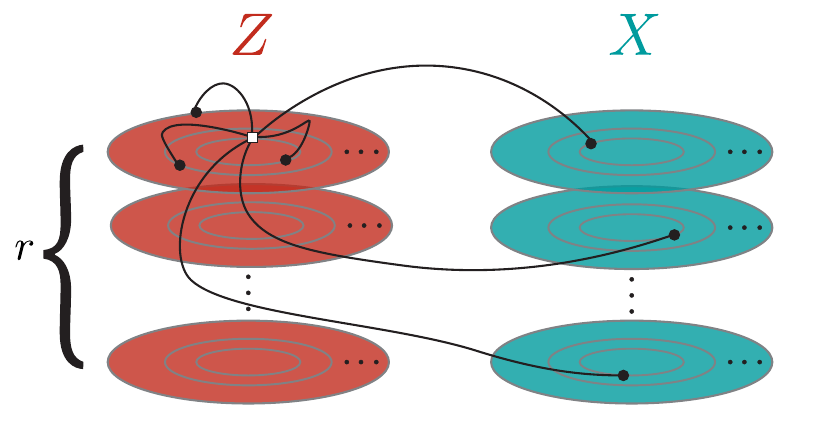}
    \caption{Representation of the structure of a quantum radial code. The code is formed from $r$ copies of a classical radial code $H_1$ (the $X$ codes) and $r$ copies of a classical radial code $H_2^*$ (the $Z$ codes). Each stabiliser is associated with a ring $u$ check in one of these classical codes and has support on a qubit from each ring of that code and a ring $u$ qubit from each code of the other type ($u=0$ in the example above)}
    \label{fig:code_stacks}
\end{figure}

\begin{figure*}[!t]
    \centering
    \includegraphics{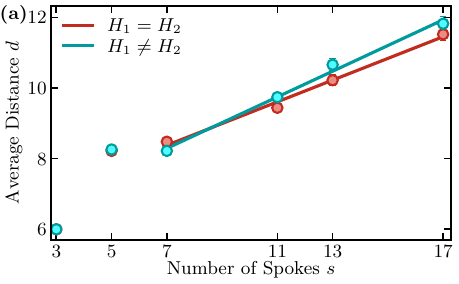}
    \includegraphics{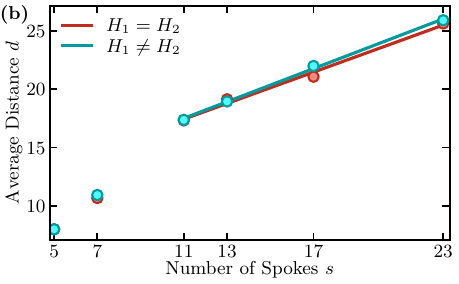}
     \caption{Numerical estimates of average distances for QRCs of different $s$ for (a) $r=3$ and (b) $r=4$. 100 codes were generated for each pair $(r,s)$. 
     Error bars are the standard error on the mean and are typically smaller than the marker.
     }
    \label{fig:av_dists}
\end{figure*}

An equivalent analysis can be performed for $Z$ stabilisers and so we obtain a direct generalisation of the structure discussed in the previous section, namely that an $X$/$Z$ stabiliser from code $c$, ring $u$ is connected to a unique qubit in each ring of code $c$ and a unique ring $u$ qubit in each $Z$/$X$ code. An illustration of this generalised structure is shown in \cref{fig:code_stacks}.

The rest of the discussion in the previous section generalises straightforwardly as well. Each codeword of a $Z$ code (all bits in rings $u'$,$u'+1$ of that code) lifts to a logical $X$ operator of the quantum code and there are $r-1$ such independent logicals per $Z$ code. The product of all $X$ stabilisers from rings $u$,$u+1$ in any $X$ code is supported on all ring $u$ and $u+1$ qubits of all $Z$ codes, so there are $r-1$ independent choices of $Z$ code and $(r-1)^2$ $X$ logicals of this type. There are also logical $X$ operators supported on all ring $u$ qubits in a pair of $X$ codes $x$ and $x+1$. There are $r$ different $X$ codes giving $(r-1)$ independent pairs of codes for each choice of ring, and the product of all $X$ stabilisers from any ring in $X$ codes $x$ and $x+1$ is supported on all qubits of these $X$ codes, meaning we have $r-1$ independent choices of ring and $(r-1)^2$ independent logicals of this type also. An identical analysis can be performed for the logical $Z$ operators, so we have $2(r-1)^2$ independent $X$ and $Z$ operators and $2(r-1)^2$ encoded qubits. The number of encoded qubits could also be checked by counting the number of relations between stabilisers, as was done in the previous example. 

Finally, we observe that every logical operator in the basis described above is supported on a pair of rings, and that every ring contains $s$ qubits, so each of these operators has weight $2s$. The number of physical qubits in the code is $2r^2s$ (as it is made of $2r$ classical radial codes, each of length $rs$) and so a general QRC has parameters $[\![2r^2s,2(r-1)^2,\leq 2s]\!]$. We were unable to identify a method to reliably achieve this upper bound on the distance, but in the next section we study the scaling of $d$ with $s$ in the average case and present numerical evidence that the relation is linear. In a similar vein, it would be preferable if the check weights could be made constant rather than scaling as $2r$, but as the majority of the arguments provided above rely on the checks of the classical codes being supported on one bit of each ring this is not possible without significantly altering the construction. However, in this work we are only concerned with small-size instances of these codes and in these cases the stabiliser weights are no more than eight.

%% file: distance.tex
\begin{figure*}[!t]
    \centering
    \includegraphics{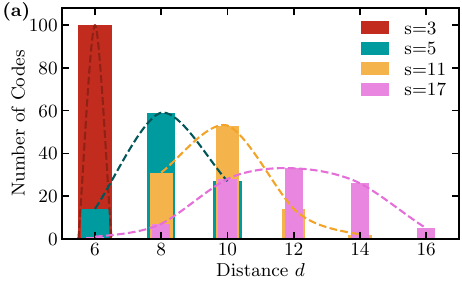}
    \includegraphics{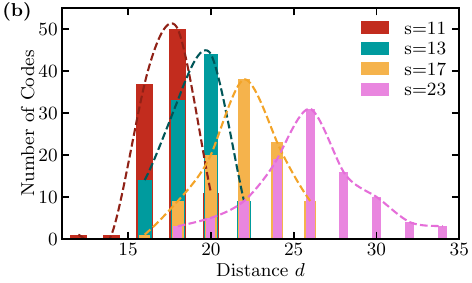}
    \caption{Distribution of distances for QRC codes of different $s$ for (a) $r=3$  and  (b) $r=4$ . The codes in these figures are a subset of the $H_1 \neq H_2$ codes from \cref{fig:av_dists}. Curves are intended to improve clarity where different distributions overlap and do not imply fitting to a model.}
    \label{fig:hists}
\end{figure*}

As shown in the example of \cref{fig:toric_code}, the upper bound on the distance of $d \leq 2s$ that we obtained previously is not generally tight, and in most cases lower distance logical operators will exist. In \cref{app:distance} we examine this specific example in more detail and conclude that QRCs obtained from products of classical codes where $H_1 = H_2$ are more likely to exhibit low-weight logicals than those where $H_1 \neq H_2$. To test this hypothesis and to get a better idea of what kinds of distances we can expect from QRCs we use QDistRnd~\cite{pryadko_qdistrnd_2022} to numerically estimate the distances of randomly generated codes for different $r$ and $s$ and for both $H_1 = H_2$ and $H_1 \neq H_2$ (i.e independently generated $H_1$ and $H_2$). 
The accuracy of these distance estimates is discussed in \cref{app:accuracy}.

We also wish to emphasise that the instances we have studied here are relatively small and there is no guarantee that this linear behaviour will continue to hold for larger codes. We note that, to the best of our knowledge, proving minimum distance bounds of lifted product codes over an Abelian group remains an open problem. Whilst certain examples of bounds exist \cite{panteleev_quantum_2022}, these rely upon the input matrices having specific structures that do not extend to the radial code construction.

Additionally, we study the distributions of these distances around the average, as shown in \cref{fig:hists}. We observe that these distances appear to be normally distributed about the mean, with the variance increasing with increasing $s$. This is unsurprising, as there is a greater variety of possible classical radial codes to use as input when $s$ is much larger than $r$. 

%% file: confinement.tex
\subsection{Confinement properties \label{ssec:confinement}}
In addition to the code distance, the confinement~\cite{quintavalle_single-shot_2021} of a code becomes relevant in the presence of syndrome measurement errors for single-shot decoding.
Formally, for a stabiliser code confinement is defined as
\begin{definition}[Confinement, \cite{quintavalle_single-shot_2021}]
    Let $t$ be an integer and $f: \mathbb{Z} \to \mathbb{R}$ some increasing function with $f(0) = 0$. We say that a stabiliser code has $(t, f)$-confinement if, for all errors $e$ with $\lvert e \rvert \leq t$, it holds
    \begin{align}
        f(\lvert \sigma(e) \lvert) \geq \lvert e \rvert^{\mathrm{red}}.
    \end{align}
\end{definition}
In the above definition, $\sigma(\bullet)$ refers to the syndrome map and the superscript $\mathrm{red}$ refers to the condition that the weight of $e$ must be reduced over the stabiliser group. Practically, we can characterise the confinement profile of a stabiliser code by searching for the minimum syndrome weights of errors of increasing weight $w\geq 1$ \cite{lin2025abelianmulticyclecodessingleshot}.
We use the software \texttt{dist-m4ri}~\cite{pryadko_dist-m4ri_2024} to compute the confinement profile of two specific examples of quantum radial codes up to $w \leq 8$ and compare it to the confinement profile of the bivariate bicycle codes in Ref.~\cite{bravyi_high-threshold_2024}. We also compare to the trivariate tricycle codes studied in Ref.~\cite{jacob_single-shot_2025}.
We restrict to $H_Z$ checks for all codes and note that trivariate tricycle codes have meta-checks in this sector.
The results are summarised in \cref{tab:confinement_profile}.

While our $\llbracket 90, 8, 10 \rrbracket$ quantum radial code has a confinement profile similar to the bivariate bicycle codes, non-decreasing only for irreducible errors of weight $w \leq 2$, our $\llbracket 352, 18, 20 \rrbracket$ code has improved confinement.
Notably, the $\llbracket 352, 18, 20 \rrbracket$ quantum radial code satiesfies $\lvert \sigma(e) \rvert \geq \lvert e \rvert^{\mathrm{red}}$ for irreducible errors with weight $w \leq 6$.
In that, it is more similar to examples of trivariate tricycle codes.
While the syndrome that an error of weight $w$ produces is larger, the condition $\lvert \sigma(e) \rvert \geq \lvert e \rvert^{\mathrm{red}}$ is satisfied for irreducible errors with weight $w \leq 7$, $w \leq 5$, and $w \leq 6$ for the three examples in \cref{tab:confinement_profile}.

\begin{figure}[!b]
    \centering
    \includegraphics[width=.5\textwidth]{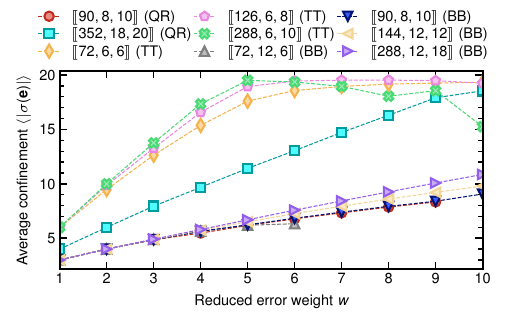}
    \caption{Average confinement profile for $X$ errors ($Z$ checks) for various codes. It represents the average syndrome weight of irreducible errors of weight $w$. Dashed lines are a guide for the eye.}
    \label{fig:average_confinement}
\end{figure}

In addition to the minimal syndrome weight for an irreducible error of weight $w$, we also show the average syndrome weight for an irreducible error of weight $w$ in \cref{fig:average_confinement}.
Note that the average syndrome weight is typically non-decreasing up to larger values of $w$ compared to the minimal syndrome weight in \cref{tab:confinement_profile}.

Lastly, it is worth noting that the confinement profile for the \textit{code} is not the same thing as the confinement profile for the \textit{detector error model} on which the decoding is performed (see \cref{section:Decoding}), although the two will be related. 
One important difference is that, in any circuit consisting of repeated measurements of the same set of stabilisers, a sequence of $w$ consecutive measurement errors on the same ancilla qubit always results in a syndrome of weight two, regardless of the size of $w$.

\begin{table}[!h]
    \centering
    \begin{tabular}{@{}cc@{}}
        \toprule
        Code & Confinement Profile \\
        \midrule
        $\llbracket 72, 12, 6\rrbracket$ (BB) 
            & 3, 4, 3, 4, 3, 4 \\
        $\llbracket 90, 8, 10 \rrbracket$ (BB) 
            & 3, 4, 3, 2, 3, 2, 3, 2 \\
        $\llbracket 108, 8, 10 \rrbracket$ (BB) 
            & 3, 4, 3, 2, 3, 2, 3, 2 \\
        $\llbracket 144, 12, 12 \rrbracket$ (BB) 
            & 3, 4, 3, 2, 3, 2, 3, 2 \\
        $\llbracket 288, 12, 18 \rrbracket$ (BB) 
            & 3, 4, 3, 2, 3, 4, 3, 4 \\
        $\llbracket 360, 12, \leq 24 \rrbracket$ (BB) 
            & 3, 4, 3, 2, 3, 4, 3, 4 \\
        \midrule
        $\llbracket 90, 8, 10 \rrbracket$ (QR) 
            & 3, 4, 3, 2, 3, 2, 3, 2 \\
        $\llbracket 352, 18, 20 \rrbracket$ (QR) 
            & 4, 6, 6, 6, 6, 6, 4, 4 \\
        \midrule
        $\llbracket 72, 6, 6\rrbracket$ (TT) 
            & 6, 8, 10, 10, 10, 10, 10, 6 \\
        $\llbracket 126, 6, 8\rrbracket$ (TT) 
            & 6, 8, 10, 12, 14, 12, 12, 6 \\
        $\llbracket 288, 6, 10\rrbracket$ (TT) 
            & 6, 8, 10, 12, 14, 14, 12, 6 \\
        \bottomrule
    \end{tabular}
    \caption{Confinement profile for $X$ errors ($Z$ checks) for various examples of bivariate bicycle~\cite{bravyi_high-threshold_2024} (BB), quantum radial (QR), and trivariate tricycle~\cite{jacob_single-shot_2025} (TT) codes.
    The confinement profile is obtained from the minimal syndrome weight of errors with weight $w = 1,\dots,8$ for all codes except the $\llbracket 72, 12, 6\rrbracket$ (BB), where $w = 1,\dots,6$.}
    \label{tab:confinement_profile}
\end{table}

%% file: decoding.tex
\subsection{Syndrome Extraction Circuits}
\label{subsection:circuits}
\begin{figure}[!b]
    \includegraphics[width=.48\textwidth]{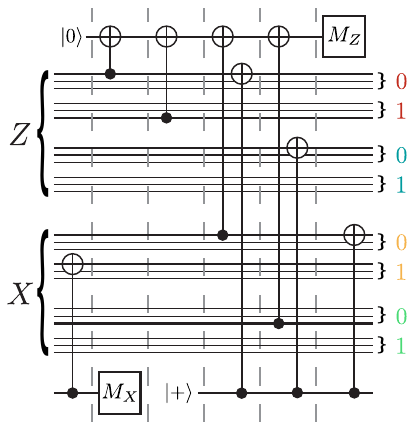}
    \caption{Measurement circuit for two stabilisers from the $(2,3)$ QRC discussed in \cref{subsection:small_example}. The top qubit is the ancilla for the ring $0$, spoke $0$ $Z$ stabiliser from the red classical code while the bottom qubit is the ancilla for the ring $0$, spoke $0$ $X$ stabiliser from the orange classical code. The 24 qubits in the middle are the 24 data qubits of the quantum code and are ordered using their associated classical codes and rings as before. Bold qubit lines are part of the support of one or both stabilisers. Dashed grey lines show boundaries between timesteps.}
    \label{fig:circuit_example}
\end{figure}

We can use the previously described structure of QRCs to define an optimal-length (in the sense that ancilla qubits are never idle) schedule for parallelised stabiliser measurements. In this schedule, each stabiliser ancilla interacts first with its associated $Z$ code qubits and then with its associated $X$ code qubits, with $X$ and $Z$ stabilisers offset from each other by half a cycle. An example for two stabilisers of the example code from \cref{subsection:small_example} is shown in \cref{fig:circuit_example}. 

The general definition of such a measurement circuit is as follows. For a $Z$ measurement ancilla associated to a stabiliser in $Z$ code $z$ and ring $u$ the associated operation at each timestep of the circuit is

\begin{itemize}
    \item $t=-1$: Initialise the ancilla in $\ket{0}$
    \item $0 \leq t < r$: Apply a CX gate with the target on the ancilla and control on the relevant qubit
    \item $r \leq t < 2r$: Apply a CX gate with the target on the ancilla and control on the relevant qubit in $X$ code $(z+t)\mod r$, ring $i$.
    \item $t=2r$: Measure the qubit in the $Z$ basis. 
\end{itemize}
For an $X$ stabiliser measurement ancilla in $X$ code $x$, ring $u$ we instead have 
\begin{itemize}
    \item $t=r-1$: Initialise the ancilla in $\ket{+}$
    \item $r \leq t < 2r$: Apply a CX gate with control on the ancilla and target on the relevant qubit in $Z$ code $(x+t)\mod r$, ring $i$
    \item $2r \leq t < 3r$: Apply a CX gate with control on the ancilla and target on the relevant qubit in $X$ code $x$, ring $(u+t)\mod r$
    \item $t=3r$: Measure the qubit in the $X$ basis.
\end{itemize}
Because each $Z$ code $z$, ring $u$ data qubit is part of a unique stabiliser from each ring of $Z$ code $z$ and a unique ring $u$ stabiliser from each $X$ code (and the reverse for $X$ code qubits) we can see that this schedule will never lead to collisions where two ancilla qubits try to interact with the same data qubit at the same time. Additionally, because each data qubit interacts first with all its associated $Z$ measurement ancillas and then with all its associated $X$ measurement ancillas the circuit is guaranteed to be valid and avoid the issues discussed in appendix B of~\cite{fowler_surface_2012}. Finally, we can see that if we run the two halves of this circuit continuously (i.e. $Z$ ancillas are initialised on each $t \mod 2r = -1$ and $X$ ancillas on each $t \mod 2r = r-1$) then these ancillas are never idle and each cycle of the circuit has the minimal possible length. 

It is worth noting that this circuit is not guaranteed to be optimal from a fault-tolerance perspective, and there may be other measurement circuits for QRCs that lead to higher effective distance. We leave the study of such circuits as a problem for future work.

\subsection{Simulation Method and Noise Model}
\label{subsection:noise}

We use \texttt{stim}~\cite{gidney_stim_2021} to simulate the propagation of noise through these measurement circuits. We use a standard noise model where for each timestep

\begin{itemize}
    \item idle qubits experience single-qubit depolarising errors with probability $p_\textrm{idle}$.
    \item pairs of qubits acted on by $CX$ gates experience two-qubit depolarising errors immediately following the gate with probability $p_\textrm{CX}$.
    \item qubits initialised in $\ket{0}$/$\ket{+}$ experience single-qubit $X/Z$ errors immediately following initialisation with probability $p_\textrm{reset}$.
    \item qubits measured in the $Z/X$ basis experience single-qubit $X/Z$ errors immediately before the measurement with probability $p_\textrm{meas}$.
\end{itemize}

We restrict our simulations to the setting where $p_\textrm{idle} = p_\textrm{CX} = p_\textrm{reset} = p_\textrm{meas}$ to facilitate comparison with other works.
However, it is worth noting that such a setting is not necessarily experimentally realistic, and that in practise e.g. idling errors may be negligible~\cite{xu_constant-overhead_2024} or that measurements error rates may be much higher than two-qubit gate error rates~\cite{gidney_fault-tolerant_2021}.

\subsection{Decoding Strategy}
\label{subsection:decoding}
The output from \texttt{stim} is a \textit{detector error model} (DEM) -- a bipartite graph of error nodes, corresponding to possible faults in the circuit, and detector nodes, corresponding to sets of measurement outcomes that can be used to infer the occurrence of these faults. An error node and a detector node are connected by an edge whenever the occurrence of the error modifies the outcome of the detector. 

Under a phenomenological noise model, the standard approach to simulating single-shot decoding is to apply random Pauli errors to the qubits of the code, calculate the corresponding syndrome, flip random bits of this syndrome to represent the action of measurement errors, and then task the decoder to calculate a correction based on this syndrome~\cite{brown_fault-tolerant_2016}. Usually, there will be some discrepancy between the calculated and true error configurations, resulting in some residual error that is carried over into the next cycle of the simulation. This approach does not generalise directly to the setting of \texttt{stim} and DEMs where error configurations are generated for the entire circuit before any decoding is performed, with some errors triggering multiple detectors with different time coordinates. To represent the action of a single-shot decoding strategy in this setting we must therefore have a way to divide the DEM into distinct time slices which can be fed independently to the decoder and to propagate the calculated corrections forward in time so that future detector outcomes are appropriately modified\footnote{Of course, it is also possible to use the measurement outcomes from \texttt{stim} to decode on the Tanner graph for the quantum code rather than on the DEM. Such a strategy may work well in codes such as the 3D gauge colour code~\cite{bombin_gauge_2015} or 3D subsystem toric code~\cite{kubica_single-shot_2022} which use matching decoders, but will work very poorly for decoders such as BP which are highly dependent on the details of the graph.}. 

\begin{algorithm}[!b]
\caption{Pseudocode for overlapping window decoding}\label{algo:owd}
\SetAlgoLined
\DontPrintSemicolon
\KwData{$\mathbf{DEM}$, $\vec{priors}$, syndrome $\vec{s}$, window $w$, commit $c$, $n_s$}
\KwResult{$\vec{total\_corr}$}

$\vec{total\_corr}$ = zeros(size(DEM, 1))\;
\For{$r$ in \textrm{range}(decodings)}{
    c\_inds, w\_inds, s\_c\_inds, s\_w\_inds = current\_inds($r$, DEM, $w$, $c$, $n_s$) \;
    \tcc{Get indices for slicing up DEM and syndrome}
    $\mathbf{DEM}_r$ = $\mathbf{DEM}$[s\_w\_inds, :]\;
    decoder = init\_decoder($\mathbf{DEM}_r$, $\vec{priors}$)\;
    $\vec{corr}$ = decoder.decode($\vec{s}$[s\_w\_inds])\:

    \If{r $\neq$ decodings-1}{
    $\vec{total\_corr}$[c\_inds] += $\vec{corr}$[c\_inds]\;
    \tcc{Propagate correction to next round}
    $\vec{s}$ += $\mathbf{DEM}$ * $\vec{total\_corr}$ \;
    }
    \Else{
    $\vec{total\_corr}$[w\_inds] += $\vec{corr}$[w\_inds]\;
    }
    \tcc{Adjust priors such that the decoder does not correct these faults again}
    $\vec{priors}$[c\_inds] = decoder.perf\_prior \;
}
\Return{$\vec{total\_corr}$}\;
\end{algorithm}

\begin{figure*}
    \includegraphics[width=.49\textwidth]{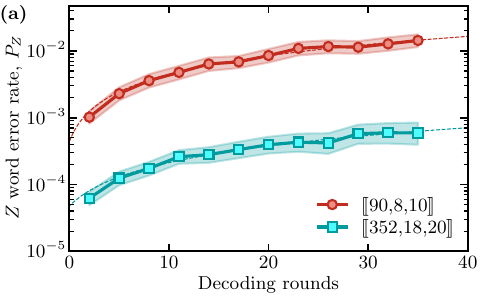}
    \includegraphics[width=.49\textwidth]{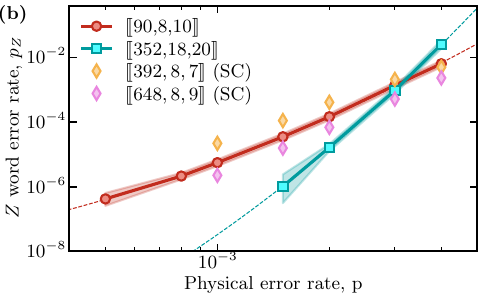}
    \caption{Word error rates in a pair of radial codes as a function of (a) number of decoding cycles and (b) physical error rate. Simulations were performed using a circuit-level noise model and  $(3,1)-$overlapping window implementation of BP-OSD-0 in (a) and BP-OSD-CS-4 in (b). In (a) the physical error rate is fixed at $p = 2 \times 10^{-3}$ and dashed lines are linear fits showing a linearly increasing error rate per decoding round after a few decoding cycles. In (b) the number of decoding cycles is fixed at $15$ and dashed lines are exponential fits with a quadratic exponent, showing exponential error suppression.
    Note that here we show the $Z$ word error rate per syndrome cycle, in contrast to (a) which showed the total $Z$ word error rate.
    Shown are also diamond markers for the $Z$ error rate of 8 patches of $d = 7$ and $d = 9$ rotated surface codes (SC) decoded over $3d$ syndrome cycles.
    Shading indicates hypotheses whose likelihoods are within a factor of 1000 of the maximum likelihood estimate.
    }
    \label{fig:main_results}
\end{figure*}

In our work, we use an overlapping window~\cite{dennis_topological_2002, berent_analog_2024, gong_toward_2024} (sometimes referred to as a sliding window~\cite{skoric_parallel_2023, huang_improved_2023}) decoding approach that fulfils these requirements.
In general, a $(w,c)-$overlapping window decoder is defined by a choice of decoding algorithm, window size $w \in \mathbb{N}$, and commit size $c \leq w \in \mathbb{N}$. The operation of the decoder is as follows. Consider a generic decoding cycle, before which a correction for the syndrome data from the first $t$ syndrome extraction cycles has already been determined. $w$ and $c$ are then used to define two sets of consecutive syndrome extraction cycles, the \textit{decoding window}, $\{t+1,t+2,...,t+w\}$, and the \textit{commit region}, $\{t+1,t+2,...,t+c\}$. In each decoding cycle, the subset of the total syndrome associated with the extraction cycles within the decoding window is identified and passed to the decoder, and a correction is returned. The parts of this correction that lie outside the commit region are then discarded and only the correction within the commit region is applied (either in hardware or software). Finally, the syndrome of this modified correction is calculated and used to update the existing syndrome data for future extraction cycles. The decoding window is then advanced by $c$ syndrome extraction cycles and the process begins again.

Alg.~\ref{algo:owd} shows an implementation of this procedure based on the DEM obtained from \texttt{stim}.
In each decoding cycle, the subgraph of the DEM lying within the decoding window is calculated using a call to $\mathrm{current\_inds}$.
We make use of the fact that in a standard stabiliser code potential faults can only trigger detectors locally, that is, a fault during a syndrome extraction cycle can only influence detectors during that and the following cycle\footnote{Note that this does not have to be true in more general codes, e.g. subsystem~\cite{higgott_subsystem_2021} and floquet/dynamic~\cite{townsend-teague_floquetifying_2023,davydova_floquet_2023} codes, where detecting regions can span multiple measurement cycles. }. %
As a result, the matrix 
representation of the DEM is a banded matrix, which allows one to determine relevant indices for the current decoding and commit region by identifying the indices of the first and last nonzero elements for a subset of rows that corresponds to the detectors of the current decoding round.

The most direct generalisation of single-shot decoding to this overlapping window case would be to take $w=c=1$, but we can also consider looser generalisations where $w$ and $c$ are simply required to be both small and independent of the size or distance of the code. We propose ``constant-depth decodable'' as an alternative and more precise term for ``single-shot decodable'' in this case.

\subsection{Numerical Results} \label{subsection:results}
For the results presented in this section, and summarised in \figref{fig:main_results}, we use the BP+OSD decoder of the \textit{ldpc} python package~\cite{roffe_decoding_2020, roffe_ldpc_2022} in combination with a $(w, c) = (3, 1)-$overlapping window decoding strategy, although a $(1, 1)$ strategy was also observed to suggest a threshold, albeit with higher word error rates.
For each window, we perform a maximum of 1000 iterations of the min-sum algorithm with a parallel schedule.
If the decoder does not converge, the soft output of the decoder is used to inform the ordered statistics subroutine.
We emphasise that these parameters are non-optimal and that further increasing, for example, the maximum number of iterations for the min-sum algorithm yields additional improvements. Examples of how logical error rates vary with the number of BP iterations are shown in \cref{app:decoder_optimization}. 
Here, we have limited simulations to 1000 iterations to obtain sufficient statistics in a reasonable time. We remark that, as its name suggests, the parallel schedule can be fully parallelised in practical applications as all necessary operations and calculations can be performed locally.
The OSD routine in its current form is not efficiently parallelisable and achieving it is an active research question~\cite{hillmann_localized_2024}.

For our simulations, we choose a pair of radial codes with parameters $\llbracket 90, 8, 10 \rrbracket$ and $\llbracket 352, 18, 20 \rrbracket$, corresponding to $(r,s)$ values of $(3, 5)$ and $(4, 11)$ respectively.
Note that the $\llbracket 90, 8, 10 \rrbracket$ code saturates the distance bound discussed in Sec.~\ref{section:distance}, while the $\llbracket 352, 18, 20 \rrbracket$ code comes close. In \sfigref{fig:main_results}{a} we can see how the $Z$ word error rates in both of these codes vary as we increase the number of simulated decoding cycles under the error model and decoding strategy described above. In these simulations, we fix the physical error rate to be $2 \times 10^{-3}$ and apply only order-0 OSD reprocessing. $X$ word error rates are not shown here but are identical within the sample variance. 
We observe that after $\sim 10$ cycles the error rate per cycle becomes constant, confirming the validity of our single-shot decoding approach. 
Further evidence for this can be seen in \sfigref{fig:main_results}{b}, which shows how the observed $Z$ word error rate per syndrome cycle $p_Z$ varies with the physical error rate $p$. 
Here we fix the number of decoding cycles to $15$ and apply order-4 reprocessing with the combination-sweep search strategy (OSD-CS-4), and observe exponential suppression of logical error rates with decreasing $p$. 
The larger code also outperforms the smaller one for low enough values of $p$. 

It is worth emphasising that the single-shot decodability of these codes is not merely a curiosity, and that the levels of error suppression displayed make them competitive with some of the best-known codes of similar size, as can be seen in the comparison against the surface code shown in \sfigref{fig:main_results}{b}.
This data for the surface code was obtained from simulations of $3d$ rounds of the  \texttt{surface\_code:rotated\_memory\_z} experiment in \texttt{stim}~\cite{gidney_stim_2021} with physical error rate $p$ for all error types. The noise model here is broadly similar to the one described above, but combines measurement and reset errors into a single error, ignores idling errors, and applies a depolarizing channel before each syndrome cycle.
The resulting error rate $P_{Z, 1}$ for a single logical qubit is then rescaled according to the expression
\begin{align}
    p_{Z} = 1 - (1 - P_{Z, 1})^{k / 3 d},
\end{align}
to obtain the $Z$ word error rate per syndrome cycle for $k = 8$ logical qubits.

Additionally, we note that in some cases the parameters and error correction capabilities of the radial codes are comparable to those of the bivariate bicycle codes~\cite{bravyi_high-threshold_2024} (e.g. when comparing the $[\![90,8,10]\!]$ codes of both constructions), but that no valid single-shot decoding strategy has been demonstrated for the latter. On the other hand, large bivariate bicycle codes appear to significantly outperform large radial codes, perhaps due to the higher check weights of the radial codes and the difference in simulated decoding strategies.

%% file: conclusion.tex
We have introduced a family of small quantum LDPC codes with high rate and distance and competitive error-suppression capabilities under a single-shot decoding strategy. We proved an exact value for $k$ and an upper bound on $d$ for all codes in this family, and showed numerically that while these codes do not achieve this bound in general, random instances can come very close to it with high probability when $n$ is suitably small. In numerical studies of performance under circuit-level noise, small examples of these codes achieved similar levels of error suppression to surface codes of comparable distance while using $\sim 5$ times fewer physical qubits. Due to their small size, flexibility and powerful error-correction capabilities, we believe these codes would be well suited to small experiments and implementations on near-term hardware, especially on e.g., neutral atom platforms~\cite{bluvstein_logical_2023} that naturally allow long-range connectivity.

A number of outstanding questions about these codes exist. 
First, we note that our results in conjunction with the confinement profile (\cref{tab:confinement_profile}) suggest that the bivariate bicycle codes of~\cite{bravyi_high-threshold_2024} could also be single-shot decodable.
Other open questions include whether more optimal measurement circuits exist for our codes -- we have not evaluated the effective distance of the circuits used in this work -- and also whether other modifications could be made to the construction to ensure higher distance.
It would also be interesting to investigate the viability of different and more efficient decoding strategies in these codes, e.g. fully local or generalised matching decoders. 

A more high-level question, given the large number of recent results involving lifted products of small classical codes, is which other classical code families might be suitable candidates for such a study and how these families might be identified. Rather than a brute-force search of the classical literature, it may be possible to identify particular desirable properties of classical codes that lead to useful quantum codes (e.g. single-shot decodability, high distance, etc).

%% file: app_pcms.tex
The following are the binary representations of the matrices \cref{eq:example_hx} and \cref{eq:example_hz} discussed in \cref{subsection:small_example}.

\begin{equation}
    H_X = 
    \resizebox{.6\textwidth}{!}{
    \begin{tikzpicture}[baseline]
        \matrix (M1) [matrix of nodes,{left delimiter=(},{right delimiter=)},ampersand replacement=\&]
        {
            1 \& 0 \& 0 \& 0 \& 0 \& 0 \& 1 \& 0 \& 0 \& 0 \& 0 \& 0 \&
            |[fill=C2]| 1 \& |[fill=C2]| 0 \& |[fill=C2]| 0 \& |[fill=C2]| 1 \& |[fill=C2]| 0 \& |[fill=C2]| 0 \& 0 \& 0 \& 0 \& 0 \& 0 \& 0 \\
            0 \& 1 \& 0 \& 0 \& 0 \& 0 \& 0 \& 1 \& 0 \& 0 \& 0 \& 0 \& 
            |[fill=C2]| 0 \& |[fill=C2]| 1 \& |[fill=C2]| 0 \& |[fill=C2]| 0 \& |[fill=C2]| 1 \& |[fill=C2]| 0 \& 0 \& 0 \& 0 \& 0 \& 0 \& 0 \\
            0 \& 0 \& 1 \& 0 \& 0 \& 0 \& 0 \& 0 \& 1 \& 0 \& 0 \& 0 \& 
            |[fill=C2]| 0 \& |[fill=C2]| 0 \& |[fill=C2]| 1 \& |[fill=C2]| 0 \& |[fill=C2]| 0 \& |[fill=C2]| 1 \& 0 \& 0 \& 0 \& 0 \& 0 \& 0 \\
            0 \& 0 \& 0 \& 1 \& 0 \& 0 \& 0 \& 0 \& 0 \& 1 \& 0 \& 0 \& 
            |[fill=C2]| 0 \& |[fill=C2]| 1 \& |[fill=C2]| 0 \& |[fill=C2]| 1 \& |[fill=C2]| 0 \& |[fill=C2]| 0 \& 0 \& 0 \& 0 \& 0 \& 0 \& 0 \\
            0 \& 0 \& 0 \& 0 \& 1 \& 0 \& 0 \& 0 \& 0 \& 0 \& 1 \& 0 \& 
            |[fill=C2]| 0 \& |[fill=C2]| 0 \& |[fill=C2]| 1 \& |[fill=C2]| 0 \& |[fill=C2]| 1 \& |[fill=C2]| 0 \& 0 \& 0 \& 0 \& 0 \& 0 \& 0 \\
            0 \& 0 \& 0 \& 0 \& 0 \& 1 \& 0 \& 0 \& 0 \& 0 \& 0 \& 1 \& 
            |[fill=C2]| 1 \& |[fill=C2]| 0 \& |[fill=C2]| 0 \& |[fill=C2]| 0 \& |[fill=C2]| 0 \& |[fill=C2]| 1 \& 0 \& 0 \& 0 \& 0 \& 0 \& 0 \\
            0 \& 1 \& 0 \& 0 \& 0 \& 0 \& 1 \& 0 \& 0 \& 0 \& 0 \& 0 \& 
            0 \& 0 \& 0 \& 0 \& 0 \& 0 \& |[fill=C3]| 1 \& |[fill=C3]| 0 \& |[fill=C3]| 0 \& |[fill=C3]| 1 \& |[fill=C3]| 0 \& |[fill=C3]| 0 \\
            0 \& 0 \& 1 \& 0 \& 0 \& 0 \& 0 \& 1 \& 0 \& 0 \& 0 \& 0 \& 
            0 \& 0 \& 0 \& 0 \& 0 \& 0 \& |[fill=C3]| 0 \& |[fill=C3]| 1 \& |[fill=C3]| 0 \& |[fill=C3]| 0 \& |[fill=C3]| 1 \& |[fill=C3]| 0 \\
            1 \& 0 \& 0 \& 0 \& 0 \& 0 \& 0 \& 0 \& 1 \& 0 \& 0 \& 0 \& 
            0 \& 0 \& 0 \& 0 \& 0 \& 0 \& |[fill=C3]| 0 \& |[fill=C3]| 0 \& |[fill=C3]| 1 \& |[fill=C3]| 0 \& |[fill=C3]| 0 \& |[fill=C3]| 1 \\
            0 \& 0 \& 0 \& 0 \& 1 \& 0 \& 0 \& 0 \& 0 \& 1 \& 0 \& 0 \& 
            0 \& 0 \& 0 \& 0 \& 0 \& 0 \& |[fill=C3]| 0 \& |[fill=C3]| 1 \& |[fill=C3]| 0 \& |[fill=C3]| 1 \& |[fill=C3]| 0 \& |[fill=C3]| 0 \\
            0 \& 0 \& 0 \& 0 \& 0 \& 1 \& 0 \& 0 \& 0 \& 0 \& 1 \& 0 \& 
            0 \& 0 \& 0 \& 0 \& 0 \& 0 \& |[fill=C3]| 0 \& |[fill=C3]| 0 \& |[fill=C3]| 1 \& |[fill=C3]| 0 \& |[fill=C3]| 1 \& |[fill=C3]| 0 \\
            0 \& 0 \& 0 \& 1 \& 0 \& 0 \& 0 \& 0 \& 0 \& 0 \& 0 \& 1 \& 
            0 \& 0 \& 0 \& 0 \& 0 \& 0 \& |[fill=C3]| 1 \& |[fill=C3]| 0 \& |[fill=C3]| 0 \& |[fill=C3]| 0 \& |[fill=C3]| 0 \& |[fill=C3]| 1 \\
        };
        \draw (M1-1-12.north east) -- (M1-12-12.south east);
        \draw [thick] (M1-1-1.north west) -- (M1-1-24.north east);
        \draw [thick] (M1-1-1.west) -- (M1-1-1.north west);
        \draw [thick] (M1-1-4.west) -- (M1-1-4.north west);
        \draw [thick] (M1-1-7.west) -- (M1-1-7.north west);
        \draw [thick] (M1-1-10.west) -- (M1-1-10.north west);
        \draw [thick] (M1-1-13.west) -- (M1-1-13.north west);
        \draw [thick] (M1-1-16.west) -- (M1-1-16.north west);
        \draw [thick] (M1-1-19.west) -- (M1-1-19.north west);
        \draw [thick] (M1-1-22.west) -- (M1-1-22.north west);
        \draw [thick] (M1-1-24.east) -- (M1-1-24.north east);
        \node [anchor=south] at (M1-1-2.north) {\color{C0} 0};
        \node [anchor=south] at (M1-1-5.north) {\color{C0} 1};
        \node [anchor=south] at (M1-1-8.north) {\color{C1} 0};
        \node [anchor=south] at (M1-1-11.north) {\color{C1} 1};
        \node [anchor=south] at (M1-1-14.north) {\color{C2} 0};
        \node [anchor=south] at (M1-1-17.north) {\color{C2} 1};
        \node [anchor=south] at (M1-1-20.north) {\color{C3} 0};
        \node [anchor=south] at (M1-1-23.north) {\color{C3} 1};
        \draw [thick] (M1-1-24.north east) -- (M1-12-24.south east);
        \draw [thick] (M1-1-24.north) -- (M1-1-24.north east);
        \draw [thick] (M1-4-24.north) -- (M1-4-24.north east);
        \draw [thick] (M1-7-24.north) -- (M1-7-24.north east);
        \draw [thick] (M1-10-24.north) -- (M1-10-24.north east);
        \draw [thick] (M1-12-24.south) -- (M1-12-24.south east);
        \node [anchor=west] at (M1-2-24.east) {\color{C2} \hspace{.5cm}0};
        \node [anchor=west] at (M1-5-24.east) {\color{C2} \hspace{.5cm}1};
        \node [anchor=west] at (M1-8-24.east) {\color{C3} \hspace{.5cm}0};
        \node [anchor=west] at (M1-11-24.east) {\color{C3} \hspace{.5cm}1};
    \end{tikzpicture}
    }
\end{equation}

\begin{equation}
    H_Z =
        \resizebox{.6\textwidth}{!}{
    \begin{tikzpicture}[baseline]
        \matrix (M1) [matrix of nodes,{left delimiter=(},{right delimiter=)},ampersand replacement=\&]
        {
            |[fill=C0!80]| 1 \& |[fill=C0!80]| 0 \& |[fill=C0!80]| 0 \& |[fill=C0!80]| 0 \& |[fill=C0!80]| 0 \& |[fill=C0!80]| 1 \& 0 \& 0 \& 0 \& 0 \& 0 \& 0 \&
            1 \& 0 \& 0 \& 0 \& 0 \& 0 \& 0 \& 0 \& 1 \& 0 \& 0 \& 0 \\ 
            |[fill=C0!80]| 0 \& |[fill=C0!80]| 1 \& |[fill=C0!80]| 0 \& |[fill=C0!80]| 1 \& |[fill=C0!80]| 0 \& |[fill=C0!80]| 0 \& 0 \& 0 \& 0 \& 0 \& 0 \& 0 \&
            0 \& 1 \& 0 \& 0 \& 0 \& 0 \& 1 \& 0 \& 0 \& 0 \& 0 \& 0 \\ 
            |[fill=C0!80]| 0 \& |[fill=C0!80]| 0 \& |[fill=C0!80]| 1 \& |[fill=C0!80]| 0 \& |[fill=C0!80]| 1 \& |[fill=C0!80]| 0 \& 0 \& 0 \& 0 \& 0 \& 0 \& 0 \&
            0 \& 0 \& 1 \& 0 \& 0 \& 0 \& 0 \& 1 \& 0 \& 0 \& 0 \& 0 \\ 
            |[fill=C0!80]| 1 \& |[fill=C0!80]| 0 \& |[fill=C0!80]| 0 \& |[fill=C0!80]| 1 \& |[fill=C0!80]| 0 \& |[fill=C0!80]| 0 \& 0 \& 0 \& 0 \& 0 \& 0 \& 0 \&
            0 \& 0 \& 0 \& 1 \& 0 \& 0 \& 0 \& 0 \& 0 \& 0 \& 0 \& 1 \\ 
            |[fill=C0!80]| 0 \& |[fill=C0!80]| 1 \& |[fill=C0!80]| 0 \& |[fill=C0!80]| 0 \& |[fill=C0!80]| 1 \& |[fill=C0!80]| 0 \& 0 \& 0 \& 0 \& 0 \& 0 \& 0 \&
            0 \& 0 \& 0 \& 0 \& 1 \& 0 \& 0 \& 0 \& 0 \& 1 \& 0 \& 0 \\ 
            |[fill=C0!80]| 0 \& |[fill=C0!80]| 0 \& |[fill=C0!80]| 1 \& |[fill=C0!80]| 0 \& |[fill=C0!80]| 0 \& |[fill=C0!80]| 1 \& 0 \& 0 \& 0 \& 0 \& 0 \& 0 \&
            0 \& 0 \& 0 \& 0 \& 0 \& 1 \& 0 \& 0 \& 0 \& 0 \& 1 \& 0 \\ 
            0 \& 0 \& 0 \& 0 \& 0 \& 0 \& |[fill=C1!50]| 1 \& |[fill=C1!50]| 0 \& |[fill=C1!50]| 0 \& |[fill=C1!50]| 0 \& |[fill=C1!50]| 0 \& |[fill=C1!50]| 1 \&
            1 \& 0 \& 0 \& 0 \& 0 \& 0 \& 1 \& 0 \& 0 \& 0 \& 0 \& 0 \\ 
            0 \& 0 \& 0 \& 0 \& 0 \& 0 \& |[fill=C1!50]| 0 \& |[fill=C1!50]| 1 \& |[fill=C1!50]| 0 \& |[fill=C1!50]| 1 \& |[fill=C1!50]| 0 \& |[fill=C1!50]| 0 \&
            0 \& 1 \& 0 \& 0 \& 0 \& 0 \& 0 \& 1 \& 0 \& 0 \& 0 \& 0 \\ 
            0 \& 0 \& 0 \& 0 \& 0 \& 0 \& |[fill=C1!50]| 0 \& |[fill=C1!50]| 0 \& |[fill=C1!50]| 1 \& |[fill=C1!50]| 0 \& |[fill=C1!50]| 1 \& |[fill=C1!50]| 0 \&
            0 \& 0 \& 1 \& 0 \& 0 \& 0 \& 0 \& 0 \& 1 \& 0 \& 0 \& 0 \\ 
            0 \& 0 \& 0 \& 0 \& 0 \& 0 \& |[fill=C1!50]| 1 \& |[fill=C1!50]| 0 \& |[fill=C1!50]| 0 \& |[fill=C1!50]| 1 \& |[fill=C1!50]| 0 \& |[fill=C1!50]| 0 \&
            0 \& 0 \& 0 \& 1 \& 0 \& 0 \& 0 \& 0 \& 0 \& 1 \& 0 \& 0 \\ 
            0 \& 0 \& 0 \& 0 \& 0 \& 0 \& |[fill=C1!50]| 0 \& |[fill=C1!50]| 1 \& |[fill=C1!50]| 0 \& |[fill=C1!50]| 0 \& |[fill=C1!50]| 1 \& |[fill=C1!50]| 0 \&
            0 \& 0 \& 0 \& 0 \& 1 \& 0 \& 0 \& 0 \& 0 \& 0 \& 1 \& 0 \\ 
            0 \& 0 \& 0 \& 0 \& 0 \& 0 \& |[fill=C1!50]| 0 \& |[fill=C1!50]| 0 \& |[fill=C1!50]| 1 \& |[fill=C1!50]| 0 \& |[fill=C1!50]| 0 \& |[fill=C1!50]| 1 \&
            0 \& 0 \& 0 \& 0 \& 0 \& 1 \& 0 \& 0 \& 0 \& 0 \& 0 \& 1 \\ 
        };
        \draw (M1-1-12.north east) -- (M1-12-12.south east);
        \draw [thick] (M1-1-1.north west) -- (M1-1-24.north east);
        \draw [thick] (M1-1-1.west) -- (M1-1-1.north west);
        \draw [thick] (M1-1-4.west) -- (M1-1-4.north west);
        \draw [thick] (M1-1-7.west) -- (M1-1-7.north west);
        \draw [thick] (M1-1-10.west) -- (M1-1-10.north west);
        \draw [thick] (M1-1-13.west) -- (M1-1-13.north west);
        \draw [thick] (M1-1-16.west) -- (M1-1-16.north west);
        \draw [thick] (M1-1-19.west) -- (M1-1-19.north west);
        \draw [thick] (M1-1-22.west) -- (M1-1-22.north west);
        \draw [thick] (M1-1-24.east) -- (M1-1-24.north east);
        \node [anchor=south] at (M1-1-2.north) {\color{C0} 0};
        \node [anchor=south] at (M1-1-5.north) {\color{C0} 1};
        \node [anchor=south] at (M1-1-8.north) {\color{C1} 0};
        \node [anchor=south] at (M1-1-11.north) {\color{C1} 1};
        \node [anchor=south] at (M1-1-14.north) {\color{C2} 0};
        \node [anchor=south] at (M1-1-17.north) {\color{C2} 1};
        \node [anchor=south] at (M1-1-20.north) {\color{C3} 0};
        \node [anchor=south] at (M1-1-23.north) {\color{C3} 1};
        \draw [thick] (M1-1-24.north east) -- (M1-12-24.south east);
        \draw [thick] (M1-1-24.north) -- (M1-1-24.north east);
        \draw [thick] (M1-4-24.north) -- (M1-4-24.north east);
        \draw [thick] (M1-7-24.north) -- (M1-7-24.north east);
        \draw [thick] (M1-10-24.north) -- (M1-10-24.north east);
        \draw [thick] (M1-12-24.south) -- (M1-12-24.south east);
        \node [anchor=west] at (M1-2-24.east) {\color{C0} \hspace{.5cm}0};
        \node [anchor=west] at (M1-5-24.east) {\color{C0} \hspace{.5cm}1};
        \node [anchor=west] at (M1-8-24.east) {\color{C1} \hspace{.5cm}0};
        \node [anchor=west] at (M1-11-24.east) {\color{C1} \hspace{.5cm}1};
    \end{tikzpicture}
    }
\end{equation}

%% file: app_lowdistance.tex
To understand what kind of features of QRCs can lead to logical operator representations with weight less than $2s$ we can return to the example discussed in \cref{subsection:small_example}, since, as we can see from \cref{fig:toric_code}, this code actually has a distance of $4$. We start by considering the product of the two $X$ stabilisers from spoke $0$ of the orange code, which can be written as a sum of binary vectors as follows

\begin{equation}
    \begin{tikzpicture}
        \matrix (M1) [matrix of nodes,ampersand replacement=\&]
        {
            1 \& 0 \& 0 \& 0 \& 0 \& 0 \& 1 \& 0 \& 0 \& 0 \& 0 \& 0 \& 1 \& 0 \& 0 \& 1 \& 0 \& 0 \& 0 \& 0 \& 0 \& 0 \& 0 \& 0 \\
            0 \& 0 \& 0 \& 1 \& 0 \& 0 \& 0 \& 0 \& 0 \& 1 \& 0 \& 0 \& 0 \& 1 \& 0 \& 1 \& 0 \& 0 \& 0 \& 0 \& 0 \& 0 \& 0 \& 0 \\
            1 \& 0 \& 0 \& 1 \& 0 \& 0 \& 1 \& 0 \& 0 \& 1 \& 0 \& 0 \& 1 \& 1 \& 0 \& 0 \& 0 \& 0 \& 0 \& 0 \& 0 \& 0 \& 0 \& 0 \\
        };
        \draw [thick] (M1-2-1.south west) -- (M1-2-24.south east);

        \node [anchor=south] at (M1-1-2.north) {\color{C0} 0};
        \node [anchor=south] at (M1-1-5.north) {\color{C0} 1};
        \node [anchor=south] at (M1-1-8.north) {\color{C0} 0};
        \node [anchor=south] at (M1-1-11.north) {\color{C0} 1};
        \node [anchor=south] at (M1-1-14.north) {\color{C2} 0};
        \node [anchor=south] at (M1-1-17.north) {\color{C2} 1};
        \node [anchor=south] at (M1-1-20.north) {\color{C3} 0};
        \node [anchor=south] at (M1-1-23.north) {\color{C3} 1};

        \node [anchor=east] at (M1-1-1.west) {\color{C2} $X(0,0)$};
        \node [anchor=east] at (M1-2-1.west) {\color{C2} $X(1,0)$};
    \end{tikzpicture}
    \label{eq:step1}
\end{equation}

We can see that when restricted to the $X$ codes the support of this operator lies within a single ring of the orange code. The product of the same two operators from the green code gives

\begin{equation}
    \begin{tikzpicture}
        \matrix (M1) [matrix of nodes,ampersand replacement=\&]
        {
            0 \& 1 \& 0 \& 0 \& 0 \& 0 \& 1 \& 0 \& 0 \& 0 \& 0 \& 0 \& 0 \& 0 \& 0 \& 0 \& 0 \& 0 \& 1 \& 0 \& 0 \& 1 \& 0 \& 0 \\
            0 \& 0 \& 0 \& 0 \& 1 \& 0 \& 0 \& 0 \& 0 \& 1 \& 0 \& 0 \& 0 \& 0 \& 0 \& 0 \& 0 \& 0 \& 0 \& 1 \& 0 \& 1 \& 0 \& 0 \\
            0 \& 1 \& 0 \& 0 \& 1 \& 0 \& 1 \& 0 \& 0 \& 1 \& 0 \& 0 \& 0 \& 0 \& 0 \& 0 \& 0 \& 0 \& 1 \& 1 \& 0 \& 0 \& 0 \& 0 \\
        };
        
        \draw [thick] (M1-2-1.south west) -- (M1-2-24.south east);

        \node [anchor=south] at (M1-1-2.north) {\color{C0} 0};
        \node [anchor=south] at (M1-1-5.north) {\color{C0} 1};
        \node [anchor=south] at (M1-1-8.north) {\color{C0} 0};
        \node [anchor=south] at (M1-1-11.north) {\color{C1} 1};
        \node [anchor=south] at (M1-1-14.north) {\color{C2} 0};
        \node [anchor=south] at (M1-1-17.north) {\color{C2} 1};
        \node [anchor=south] at (M1-1-20.north) {\color{C3} 0};
        \node [anchor=south] at (M1-1-23.north) {\color{C3} 1};

        \node [anchor=east] at (M1-1-1.west) {\color{C3} $X(0,0)$};
        \node [anchor=east] at (M1-2-1.west) {\color{C3} $X(1,0)$};
    \end{tikzpicture}
    \label{eq:step2}
\end{equation}

\noindent whose restriction to the $X$ codes lies within a single ring of the green code. The product of these two stabilisers is then

\begin{equation}
    \begin{tikzpicture}
        \matrix (M1) [matrix of nodes,ampersand replacement=\&]
        {
            1 \& 0 \& 0 \& 1 \& 0 \& 0 \& 1 \& 0 \& 0 \& 1 \& 0 \& 0 \& 1 \& 1 \& 0 \& 0 \& 0 \& 0 \& 0 \& 0 \& 0 \& 0 \& 0 \& 0 \\
            0 \& 1 \& 0 \& 0 \& 1 \& 0 \& 1 \& 0 \& 0 \& 1 \& 0 \& 0 \& 0 \& 0 \& 0 \& 0 \& 0 \& 0 \& 1 \& 1 \& 0 \& 0 \& 0 \& 0 \\
            1 \& 1 \& 0 \& 1 \& 1 \& 0 \& 0 \& 0 \& 0 \& 0 \& 0 \& 0 \& 1 \& 1 \& 0 \& 0 \& 0 \& 0 \& 1 \& 1 \& 0 \& 0 \& 0 \& 0 \\
        };
        
        \draw [thick] (M1-2-1.south west) -- (M1-2-24.south east);

        \node [anchor=south] at (M1-1-2.north) {\color{C0} 0};
        \node [anchor=south] at (M1-1-5.north) {\color{C0} 1};
        \node [anchor=south] at (M1-1-8.north) {\color{C1} 0};
        \node [anchor=south] at (M1-1-11.north) {\color{C1} 1};
        \node [anchor=south] at (M1-1-14.north) {\color{C2} 0};
        \node [anchor=south] at (M1-1-17.north) {\color{C2} 1};
        \node [anchor=south] at (M1-1-20.north) {\color{C3} 0};
        \node [anchor=south] at (M1-1-23.north) {\color{C3} 1};

        \node [anchor=east] at (M1-1-1.west) {\color{C2} $X(0,0)+X(1,0)$};
        \node [anchor=east] at (M1-2-1.west) {\color{C3} $X(0,0)+X(1,0)$};
    \end{tikzpicture}
    \label{eq:step3}
\end{equation}

\noindent which, when restricted to the $Z$ codes, is only supported on the red code. Recall that there is a logical $X$ operator supported on all qubits of rings $0$ and $1$ in the red code and another operator supported on all qubits of ring $0$ in the orange and green codes. The product of these two operators with the stabiliser shown above is then  

\begin{equation}
    \begin{tikzpicture}
        \matrix (M1) [matrix of nodes,ampersand replacement=\&]
        {
            1 \& 1 \& 1 \& 1 \& 1 \& 1 \& 0 \& 0 \& 0 \& 0 \& 0 \& 0 \& 1 \& 1 \& 1 \& 0 \& 0 \& 0 \& 1 \& 1 \& 1 \& 0 \& 0 \& 0 \\
            1 \& 1 \& 0 \& 1 \& 1 \& 0 \& 0 \& 0 \& 0 \& 0 \& 0 \& 0 \& 1 \& 1 \& 0 \& 0 \& 0 \& 0 \& 1 \& 1 \& 0 \& 0 \& 0 \& 0 \\
            0 \& 0 \& 1 \& 0 \& 0 \& 1 \& 0 \& 0 \& 0 \& 0 \& 0 \& 0 \& 0 \& 0 \& 1 \& 0 \& 0 \& 0 \& 0 \& 0 \& 1 \& 0 \& 0 \& 0 \\
        };
        
        \draw [thick] (M1-2-1.south west) -- (M1-2-24.south east);

        \node [anchor=south] at (M1-1-2.north) {\color{C0} 0};
        \node [anchor=south] at (M1-1-5.north) {\color{C0} 1};
        \node [anchor=south] at (M1-1-8.north) {\color{C1} 0};
        \node [anchor=south] at (M1-1-11.north) {\color{C1} 1};
        \node [anchor=south] at (M1-1-14.north) {\color{C2} 0};
        \node [anchor=south] at (M1-1-17.north) {\color{C2} 1};
        \node [anchor=south] at (M1-1-20.north) {\color{C3} 0};
        \node [anchor=south] at (M1-1-23.north) {\color{C3} 1};

        \node [anchor=east] at (M1-1-1.west) {$\overline{X_1} + \overline{X_2}$};
        \node [anchor=east] at (M1-2-1.west) {$X_\mathrm{stab}$};
    \end{tikzpicture}
    \label{eq:step4}
\end{equation}

\noindent which only has weight 4. 

Our ability to reduce the weight of the logical operator in this way depended on three things. (a) It was possible to take products of $X$ stabilisers associated with a single $X$ code in such a way that the resulting stabiliser was supported on only a single ring of this $X$ code (when restricted to the $X$ codes). (b) It was possible to take products of $X$ stabilisers associated with different $X$ codes in such a way that the resulting stabiliser was supported on only a single $Z$ code (when restricted to the $Z$ codes). (c) We were able to do both (a) and (b) simultaneously. 

In particular, the ability to achieve (c) came from the fact that $H_1 = H_2$ and so the structure of the $X$ stabilisers within $X$ codes ($H_2$) and between $X$ and $Z$ codes ($H_1$) were the same. This fact meant that in \cref{eq:step3} taking a product of stabilisers restricted to a single ring in the $X$ codes naturally gave rise to a stabiliser restricted to a single code in the $Z$ codes. Choosing $H_1$ and $H_2$ to be very different means that a product of generators that produces a very structured operator in the $X$ codes will generally produce a very unstructured operator in the $Z$ codes and vice versa. 

Numerical investigation (\cref{fig:av_dists}) suggests that choosing $H_1 \neq H_2$ does lead to an increase in average-case code distance in some regimes, but this increase is fairly small. Methods for reducing the likelihood of (a) and (b) were also developed but were not observed to lead to any distance improvements and so are not described here. An interesting question for future research is why the lifted product can so drastically reduce the distance of the input codes, and how this can (as much as possible) be prevented.

%% file: app_accuracy.tex
As described in~\cite{pryadko_qdistrnd_2022}, the probability $P_\textrm{fail}$ that the distance $\Tilde{d}$ calculated by QDistRnd is greater than the true code distance is expected to be upper bounded by

\begin{equation}
    P_\textrm{fail} < e^{-\langle n \rangle}
\end{equation}

\noindent where $\langle n \rangle$ is the average number of times a codeword of weight $\Tilde{d}$ was found.

\begin{center}
    \includegraphics{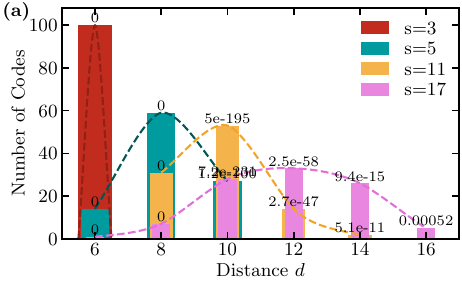}
    \includegraphics{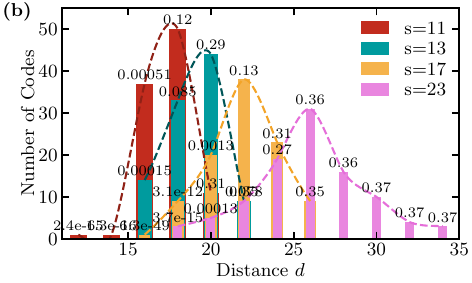}
\end{center}

\begin{figure*}
    \centering
    \includegraphics{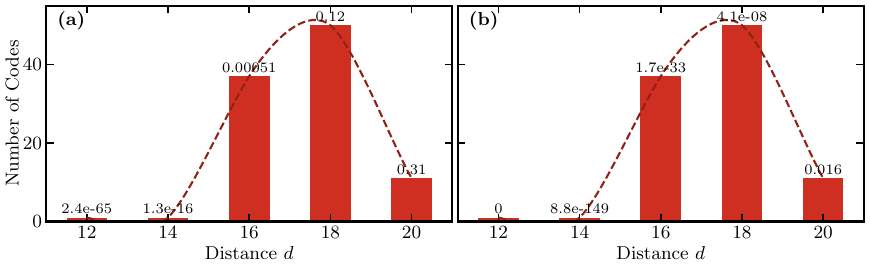}
    \caption{Distance distributions for the $(4, 11)$ quantum radial code obtained from QDistRand for an increasing number of samples, i.e., $10^5$ in (a) and $10^6$ in (b).
    Annotated numbers on each bar indicate the upper bound on the probability that the distance $\Tilde{d}$ calculated by QDistRand is greater than the true code distance.}
    \label{fig:qdist_rand_dist_estimates}
\end{figure*}

Above we show versions of the plots from \cref{fig:hists} with the average upper bound on failure probability displayed for each $(r,s,\Tilde{d})$. We can see that for $r=3$ the failure probability is negligible for all except the highest distance codes. For $r=4$, on the other hand, this bound is much higher, on the order of $0.1$ for almost all codes. However, we believe that these distance estimates are reliable for the following reasons: Firstly, they are consistent with the pattern of distance distributions observed for the $r=3$ codes, where the bound on failure probability is much lower. Secondly, the distance estimates are symmetrically distributed about the mean and failure probabilities for codes at the low end of this distribution are fairly small. As there is no obvious reason why the distribution should be asymmetric (especially as all $r=3$ distributions are symmetric) this suggests that the upper end of the distribution also has roughly the correct shape. Finally, increasing the number of samples in QDistRnd does not noticably affect the distribution shapes. For instance, \Cref{fig:qdist_rand_dist_estimates} shows the distributions of $(4,11)$ codes with $10^5$ (left) and $10^6$ (right) samples.

The distributions are identical in both cases but the bounds on the failure probabilities of the latter are much lower than the former. We conjecture that, in the same way, taking additional samples would reduce the bounds on failure probabilites for all distibutions in \cref{fig:hists} without noticably altering distribution shapes. Unfortunately, the number of samples and runtime required to test this for higher values of $s$ would be prohibitively large.

%% file: app_code_defs.tex
The $[\![90,8,10]\!]$ code studied numerically in \cref{section:Decoding} is the lifted product of the classical codes defined by 

\begin{equation}
    A_1 = 
    \begin{pmatrix}
        3 & 2 & 1 \\
        4 & 1 & 4 \\
        1 & 2 & 3 
    \end{pmatrix}_5
    ~~
    A_2 = 
    \begin{pmatrix}
        3 & 3 & 0 \\
        1 & 0 & 1 \\
        4 & 2 & 0
    \end{pmatrix}_5
\end{equation}

\noindent and for the $[\![352,18,20]\!]$ code

\begin{equation}
    A_1 = 
    \begin{pmatrix}
        10 & 10 & 1 & 6 \\
        4 & 7 & 5 & 2 \\
        8 & 10 & 6 & 9 \\
        1 & 6 & 0 & 6
    \end{pmatrix}_{11}
    ~~
    A_2 = 
    \begin{pmatrix}
        9 & 5 & 8 & 3 \\
        5 & 4 & 1 & 0 \\
        0 & 4 & 6 & 10 \\
        2 & 8 & 4 & 2
    \end{pmatrix}_{11}
\end{equation}

%% file: app_decoder_optimization.tex
\begin{figure*}
    \centering
    \includegraphics[width=0.49\textwidth]{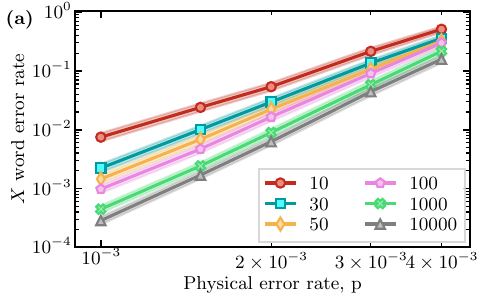}
    \includegraphics[width=0.49\textwidth]{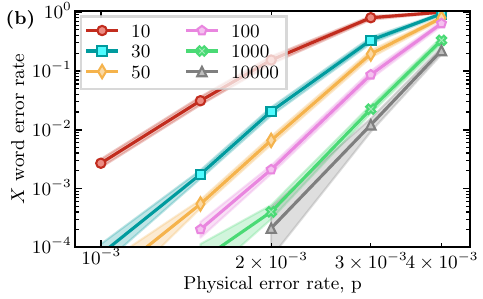}
    \caption{Logical $X$ error rate after 15 decoding rounds with the $(3, 1)$-overlapping window decoder in (a)  for the $\llbracket 90, 8, 10 \rrbracket$ and in (b) for $\llbracket 352, 18, 20 \rrbracket$ code. Here we varied the number of maximum iterations in the min-sum decoder, shown in the legend, and fixed the reprocessing routine to OSD-0.
    A very large number of iterations are necessary before performance begins to plateau.}
    \label{fig:bp_iteration_sweep}
\end{figure*}

It has been suggested that the performance of BP+OSD in surface codes is limited by the existence of short loops in the decoding graph~\cite{roffe_decoding_2020,higgott_improved_2023}, which prevent meaningful propagation of information once the iteration count becomes roughly equal to this loop length. However, such loops also exist in the decoding graphs of the codes described in this work, and indeed, in all decoding graphs for stabiliser codes which use repeated syndrome measurements~\cite[see App. B]{berent_analog_2024}, and here we observe continued and substantial improvement in the performance of BP up to very high numbers of iterations ($\sim 10000$, see \cref{fig:bp_iteration_sweep}). We conjecture that the poor performance observed for BP in surface codes is due not just to the existence of short loops but to the combination of both short loops and pointlike syndromes.

The results in appendix B of~\cite{higgott_improved_2023} suggest that the relevant part of the BP algorithm is the communication between factor nodes with syndrome $-1$ (this makes intuitive sense as these are the non-trivial features of the syndrome). We can think of each pair of these nodes (A, B) as being connected by a noisy channel, with the noise strength proportional to the separation distance between the nodes. Once this separation becomes roughly equal to the girth $g$ of the graph, the channel becomes effectively completely noisy, preventing meaningful communication. However, if there exists a sequence of node pairs $(A, B), (B, C), \dots (Y, Z)$ where the distance between consecutive nodes in each pair is less than $g$, information can still propagate through this path, albeit very slowly.

In the case of pointlike syndromes such paths are unlikely to exist, but they can occur very naturally in codes such as those studied here, where syndromes are extensive (i.e. the syndrome weight is non-constant in the error weight). This slow propagation of information across long, interconnected paths in codes with extensive syndromes potentially explains why a higher number of BP iterations can significantly improve error rates in these codes.

